\documentclass{ws-ijmpa}

\usepackage[super]{cite}
\usepackage[dvipsnames]{xcolor}
\usepackage[verbose,hypertexnames=false]{hyperref}
\usepackage{tikz}
\hypersetup{colorlinks=false,allbordercolors=blue,pdfborderstyle={/S/U/W 1}}

\begin{document}

\markboth{Kunio Kaneta, Hye-Sung Lee, Jiheon Lee, and Jaeok Yi}{Dark Energy under a Gauge Symmetry: A Review of Gauged Quintessence and Its Implications}

\catchline{}{}{}{}{}

\title{Dark energy under a gauge symmetry:\\ A review of gauged quintessence and its implications}

\author{Kunio Kaneta}
\address{Faculty of Education, Niigata University, Niigata 950-2181, Japan\\
kaneta@ed.niigata-u.ac.jp}

\author{Hye-Sung Lee}
\address{Department of Physics, KAIST, Daejeon 34141, Korea\\
hyesung.lee@kaist.ac.kr}

\author{Jiheon Lee}
\address{Department of Physics, KAIST, Daejeon 34141, Korea\\
anffl0101@kaist.ac.kr}

\author{Jaeok Yi}
\address{Department of Physics, KAIST, Daejeon 34141, Korea\\
wodhr1541@kaist.ac.kr}

\maketitle

\begin{history}
\today
\end{history}

\begin{abstract}
We review the \emph{gauged quintessence} scenario, wherein the quintessence scalar field responsible for dark energy is promoted to a complex field charged under a dark $U(1)$ gauge symmetry. This construction leads to new and potentially rich cosmological phenomenology. After a concise recap of the standard quintessence scenario, we highlight how a $U(1)$ gauge invariance alters the dynamics of the scalar and the associated dark gauge boson. We survey the evolution of both fields across cosmic history, discuss their possible production via a misalignment mechanism, and examine implications for the Hubble tension. We also comment on potential non-gravitational signals of gauged quintessence through kinetic mixing (the dark photon vector portal). 
\end{abstract}


\tableofcontents

\section{Introduction}
\label{sec:intro}
Over the past few decades, our understanding of the universe has converged on a cosmological model in which approximately 5\% of the total energy budget is composed of ordinary (baryonic) matter, 27\% is dark matter, and the remaining 68\% is dark energy. The ordinary matter sector is well-described by the Standard Model (SM) of particle physics, which is based on the gauge group \(SU(3) \times SU(2) \times U(1)\). By contrast, dark matter (DM) remains elusive despite extensive research efforts through direct detection experiments, indirect searches (e.g., from astrophysical signals), and collider-based probes. Numerous theories propose additional symmetries in the dark sector, such as a hidden \(U(1)\) gauge symmetry with kinetic mixing to the SM hypercharge\cite{Holdom:1985ag}, a Peccei--Quinn-type \(U(1)_{PQ}\) for axion dark matter\cite{Peccei:1977hh,Wilczek:1977pj,Weinberg:1977ma,Kim:1979if,Shifman:1979if,Zhitnitsky:1980tq,Dine:1981rt}, or supersymmetric extensions\cite{Wess:1974tw} in which dark matter candidates interact weakly through the SM gauge group.

Dark energy, on the other hand, is arguably the least understood component of the cosmological energy budget. Its leading observational signature is the accelerated expansion of the universe, which is studied through cosmological measurements of expansion rates at different redshifts.\footnote{Interactions of dark energy with baryons have also been proposed to influence cosmic evolution~\cite{Vagnozzi:2019kvw,Ferlito:2022mok}, and screened interactions between dark energy and photons might explain anomalies such as the one observed by XENON1T~\cite{Vagnozzi:2021quy}.} However, unlike the dark matter sector, there is no widely accepted theoretical framework that incorporates a well-motivated non-gravitational symmetry for dark energy. This gap has limited the scope of systematic symmetry-based approaches in dark energy research.

To address this issue, we have introduced a model termed \emph{gauged quintessence} \cite{Kaneta:2022kjj,Kaneta:2023lki,Kaneta:2023wdr}. Quintessence~\cite{Ratra:1987rm,Peebles:1987ek,Caldwell:1997ii} postulates a singlet scalar field that slowly evolves over cosmological timescales, acting as a dynamical source of dark energy, which has diversified into numerous distinct variations~\cite{Frieman:1995pm,Carroll:1998zi,Kim:1998kx,Choi:1999xn,Gu:2001tr,Brisudova:2001wb,Boyle:2001du,Li:2001xaa,Brisudova:2001ur,Hill:2002kq,Kim:2002tq,Mainini:2004he,Rinaldi:2014yta,Mehrabi:2015lfa,Alvarez:2019ues,Orjuela-Quintana:2020klr,Motoa-Manzano:2020mwe}. Our gauged quintessence scenario extends the original idea by incorporating a gauge symmetry into the quintessence field itself, thereby offering new avenues for theoretical exploration and potential observational signatures through the interaction of dark energy with other sectors.

The remainder of this review is organized as follows. In Section~\ref{sec:quintessence}, we provide a brief overview of the conventional quintessence framework. Section~\ref{sec:GQ} presents the construction of the gauged quintessence model, highlighting its novel features.
The quantum corrections are discussed in Section~\ref{sec:quantum_corrections}.
In Section~\ref{sec:misalignment}, we discuss the misalignment mechanism and resulting vector boson production within this framework.
Section~\ref{sec:evolution} provides an analysis of how gauged quintessence may affect the evolution of the universe.
We then explore possible implications of gauged quintessence for the Hubble tension in Section~\ref{sec:hubble_tension}. Finally, in Section~\ref{sec:signals}, we briefly comment on potential non-gravitational signals arising from dark energy under a gauge symmetry. 

The goal of this review is to provide a clear perspective on how an additional gauge structure in the dark energy sector can not only inform theoretical modeling but also offer novel observational possibilities. We hope this work will motivate further investigations into the symmetry principles that may underlie the dynamics of dark energy.

\section{Quintessence Basics}
\label{sec:quintessence}
\begin{figure}[bt]
\centerline{\includegraphics[width=0.8\linewidth]{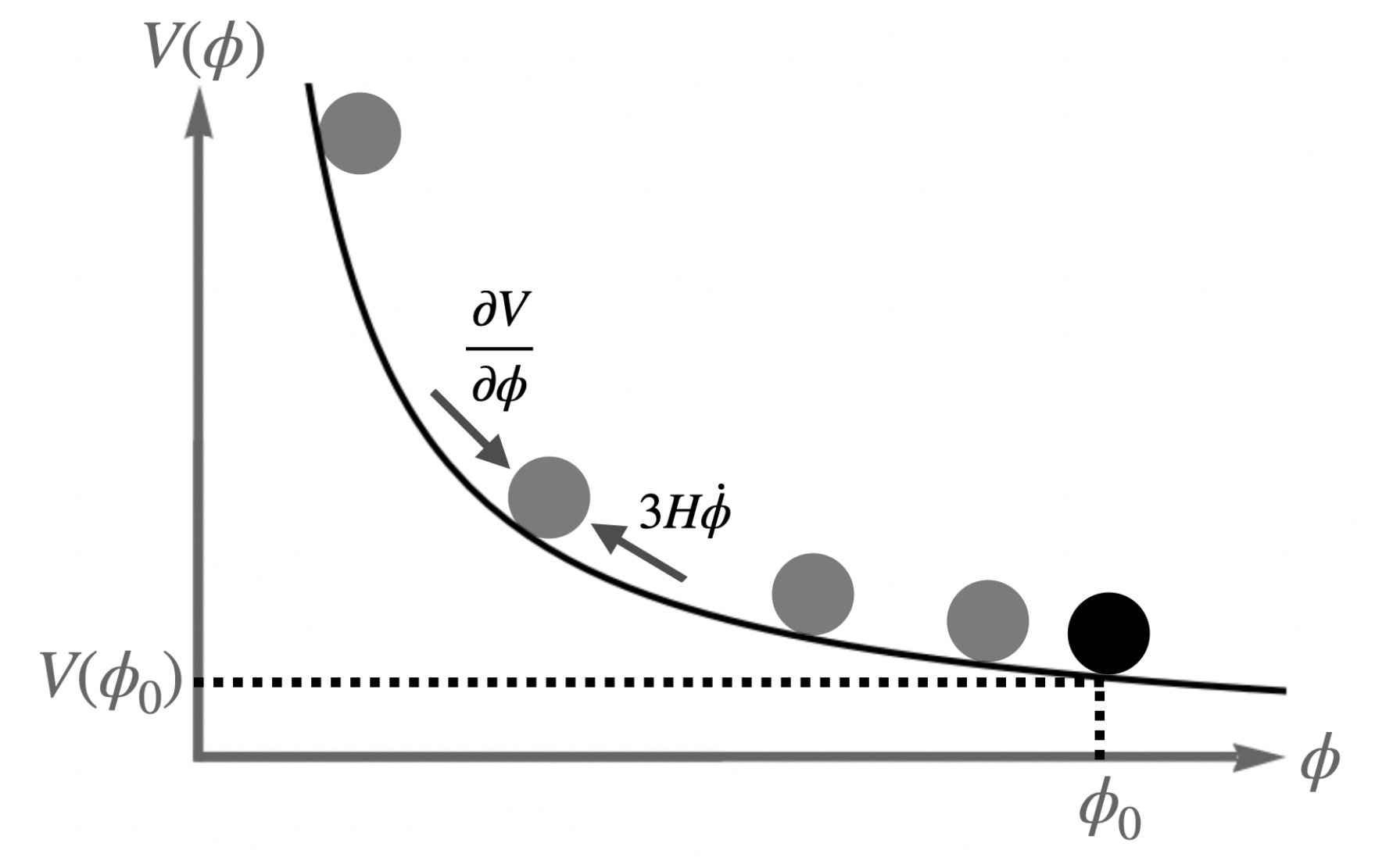}}
\caption{A schematic illustration of a runaway quintessence potential. As \(\phi\) slowly rolls down this potential, its potential energy can remain sufficiently large to drive accelerated cosmic expansion.}
\label{fig:runaway}
\end{figure}

In this section, we briefly review the standard (ungauged) quintessence framework. Quintessence is a dynamical dark energy model in which a scalar field \(\phi\) evolves slowly at late times, such that its potential energy drives the accelerated expansion of the Universe (see Fig.~\ref{fig:runaway}). Historically, Ratra and Peebles introduced two representative potentials for quintessence: an inverse power-law potential and an exponential potentials \cite{Ratra:1987rm,Peebles:1987ek}. The former takes the form
\begin{equation}
    V_0(\phi) = \frac{M^{\alpha+4}}{|\phi|^\alpha},
    \label{eq:V0}
\end{equation}
where \(\alpha>0\) and \(M\) is a parameter with mass dimension. 
This kind of potential is often characterized as a ``runaway'' potential, unbounded from below, and is phenomenological rather than a UV-complete description.

An important feature of quintessence is the \emph{tracking behavior}, first emphasized by Steinhardt, Wang, and Zlatev \cite{Steinhardt:1999nw}. In a tracking solution, the present-day value of \(\phi\) is largely insensitive to its initial conditions, thereby alleviating the cosmological coincidence problem. This occurs because the field’s evolution is largely determined by the background energy density. Since dark energy and matter scale differently with the cosmic scale factor \(a(t)\) (matter density scales as \(a^{-3}\) while the evolution of dark energy is not strictly restricted except it remains nearly constant at present), it would otherwise be difficult to arrange for their densities to be comparable in the present epoch without fine-tuning. Tracking solutions help mitigate this fine-tuning, independent of the initial condition of the scalar field.

\subsection{Action and Equations of Motion}
\label{subsec:quintessence-action}

The simplest quintessence model introduces a real scalar field \(\phi\) with a canonical kinetic term and a potential \(V_0(\phi)\). In a flat Friedmann–Lema\^itre–Robertson–Walker (FLRW) universe with metric
\(
g_{\mu\nu} = \mathrm{Diag}\{-1,\,a(t)^2,\,a(t)^2,\,a(t)^2\},
\)
the relevant action can be written as
\begin{equation}
    S_\mathrm{Q} 
    =
    \int d^4x \,\sqrt{-g}\,\Bigg[\frac{1}{2} m_{\mathrm{Pl}}^2 R 
    \;-\; \frac{1}{2} g^{\mu\nu}\,\partial_\mu\phi\,\partial_\nu\phi 
    \;-\; V_0(\phi)\Bigg],
    \label{eq:QuintessenceAction}
\end{equation}
where \(m_{\text{Pl}} = M_{\text{Pl}}/\sqrt{8\pi} \approx 2.4\times10^{18}\,\mathrm{GeV}\) is the reduced Planck mass, \(M_{\text{Pl}} \approx 1.2\times10^{19}\,\mathrm{GeV}\) is the Planck mass.
\(R\) is the Ricci scalar, and \(g = \det(g_{\mu\nu})\).

Assuming spatial homogeneity so that \(\phi = \phi(t)\), one obtains the Klein-Gordon equation of motion:
\begin{equation}
    \ddot{\phi}+ 3H\dot{\phi} 
    + \frac{dV_0}{d\phi}
    = 0,
    \label{eq:KG}
\end{equation}
where \(H(t) = \dot{a}(t)/a(t)\) is the Hubble parameter, and the term \(3H\dot{\phi}\) provides a friction-like effect that can significantly slow the scalar field’s evolution.

The energy density \(\rho_{\phi}\) and pressure \(p_{\phi}\) of the quintessence field are given by
\begin{equation}
    \rho_{\phi} 
    =
    \frac{1}{2}\,\dot{\phi}^2 
    +
    V_0(\phi),
    \qquad
    p_{\phi}
    =
    \frac{1}{2}\,\dot{\phi}^2 
    -
    V_0(\phi).
    \label{eq:rho_phi & p_phi}
\end{equation}
Their ratio defines the equation of state parameter,
\begin{equation}
    w_{\phi} 
    \;=\; 
    \frac{p_{\phi}}{\rho_{\phi}} 
    \;=\; 
    \frac{\frac{1}{2}\dot{\phi}^2 - V_0(\phi)}
         {\frac{1}{2}\dot{\phi}^2 + V_0(\phi)},
    \label{eq:eos_quint}
\end{equation}
which represents how the energy density of quintessence evolves with cosmic expansion. 
For quintessence to drive cosmic acceleration at late times, one generally requires \(w_{\phi} < -\frac{1}{3}\). In the slow-roll limit, where \(\dot{\phi}^2 \ll V_0(\phi)\), this parameter approaches \(w_{\phi} \simeq -1\), which is the cosmological constant $\Lambda$ limit. Due to the positivity of the kinetic energy, the equation of state for quintessence cannot be less than $-1$, i.e., \(w_{\psi} \geq -1 \).

\subsection{Mass Scale and Present-Day Conditions}
\label{subsec:quintessence-mass}

Two crucial conditions are required for quintessence to account for the observed dark energy:
\begin{enumerate}
    \item The scalar potential must match the dark energy scale today, i.e.,
    \begin{equation}
        V_0(\phi_0)
        \;\simeq\; 
        10^{-123}\,m_{\mathrm{Pl}}^4 
        \;\sim\; 
        3\times10^{-47}\,\mathrm{GeV}^4.
    \end{equation}
    This corresponds to the current dark energy density determined by the observation\cite{Planck:2018vyg}.

    \item The effective mass of the quintessence field,
    \begin{equation}
        m_{\phi}^2 
        \;=\; 
        \frac{\partial^2 V_0}{\partial \phi^2},
        \label{eq:mass}
    \end{equation}
    should be comparable to or less than the present-day Hubble parameter, 
    \begin{equation}
        m_{\phi} 
        \;\lesssim\; 
        H_0 
        \;\sim\; 
        10^{-42}\,\mathrm{GeV}.
    \end{equation}
    This ensures that the friction term \(3H\dot{\phi}\) remains significant, keeping the field in slow-roll evolution.
\end{enumerate}

Because \(H_0\) (associated with the age of the Universe) represents one of the lowest energy scales in nature, it follows that the quintessence mass must be extremely light. This requirement underlines the difficulty of embedding quintessence in high-energy frameworks, such as supersymmetric models, while avoiding undesirable corrections that could destabilize such a light scalar field~\cite{Kolda:1998wq,Binetruy:1998rz,Brax:1999gp}. Nevertheless, the basic quintessence picture offers a compelling explanation for the observed accelerated expansion and continues to serve as a key benchmark against which alternative or extended dark energy models are compared.

\section{Gauged Quintessence}
\label{sec:GQ}
We now extend the standard quintessence framework by introducing a dark \(U(1)\) gauge symmetry. In this construction, the quintessence field is promoted to a complex scalar \(\Phi\), whose radial component \(\phi\) plays the role of the quintessence (or ``dark Higgs'') field, while its phase \(\eta\) is the longitudinal component of the associated dark gauge boson \(X_\mu\). Consequently, this dark Higgs mechanism endows the gauge boson with a mass that varies as \(\phi\) evolves cosmologically.

\subsection{Action and Gauge Invariance}
\label{subsec:GQ-action}

The full action takes the form
\begin{equation}
S = \int d^4 x \,\sqrt{-g} \,\biggl[
\frac{1}{2} m_{\text{Pl}}^{2} \,R 
- \bigl|D_{\mu}\Phi\bigr|^{2} 
- V_{0}(\Phi) 
- \frac{1}{4}\,\mathbb{X}_{\mu\nu}\,\mathbb{X}^{\mu\nu}
\biggr].
\label{eq:action}
\end{equation}
The covariant derivative is defined as
\begin{equation}
D_\mu \equiv \partial_\mu+ i\,g_X \mathbb{X}_\mu,
\end{equation}
with \(g_X\) being the dark gauge coupling constant, and 
\(\mathbb{X}_{\mu\nu} \equiv \partial_\mu \mathbb{X}_\nu - \partial_\nu \mathbb{X}_\mu\). 

For simplicity, we impose the unitary gauge, in which the phase \(\eta\) is absorbed into the gauge field:
\begin{equation}
\eta = 0, 
\quad\quad
X_{\mu} = \mathbb{X}_{\mu} + \frac{1}{g_X}\,\partial_{\mu}\eta.
\end{equation}
In this gauge, the complex scalar \(\Phi\) reduces to a real scalar field \(\phi\), and the action becomes
\begin{equation}
S = \int d^4 x \,\sqrt{-g}\biggl[
\frac{1}{2}m_{\text{Pl}}^2 R 
- \frac{1}{2}\bigl(\partial_{\mu}\phi\bigr)^{2} 
- \frac{1}{4}X_{\mu\nu}X^{\mu\nu}
- V_0(\phi) 
- \frac{1}{2}\,g_X^{2}\phi^{2}X_{\mu}X^{\mu}
\biggr].
\end{equation}
Note that \(X_{\mu\nu} = \partial_\mu X_\nu - \partial_\nu X_\mu\) in this gauge. The term
\begin{equation}
V_{\text{gauge}}(\phi) 
= \frac{1}{2}g_X^2\phi^2X_{\mu}X^{\mu}
\end{equation}
represents the \emph{gauge potential}, which endows the dark gauge boson \(X\) with a mass \(m_X^2 = g_X^2 \phi^2\). In this sense, \(\phi\) serves as a dark Higgs field, and $V_{\text{gauge}}(\phi)$ plays a very crucial role.

\subsection{Mass Evolution and Boltzmann Equations}
\label{subsec:GQ-mass}

In addition to the original quintessence potential \(V_0(\phi)\), taken here to be the Ratra--Peebles inverse-power potential in Eq.~\eqref{eq:V0}, the gauge potential modifies the effective mass of both \(\phi\) and \(X\). At tree level, the mass terms are given by
\begin{equation}
m_{\phi}^2\bigl|_{0} 
\;=\; \frac{\partial^2 V_0}{\partial \phi^2} 
\;+\; \frac{\partial^2 V_{\text{gauge}}}{\partial \phi^2},
\quad\quad
m_{X}^2\bigl|_{0} 
\;=\; g_X^{2}\,\phi^{2}.
\label{eq:treemass}
\end{equation}
Here \((\cdot)|_{0}\) indicates evaluation at the \(\phi\) value determined by the (instantaneous) tree-level potential.

\begin{figure}[tb]
\centerline{\includegraphics[width=0.8\linewidth]{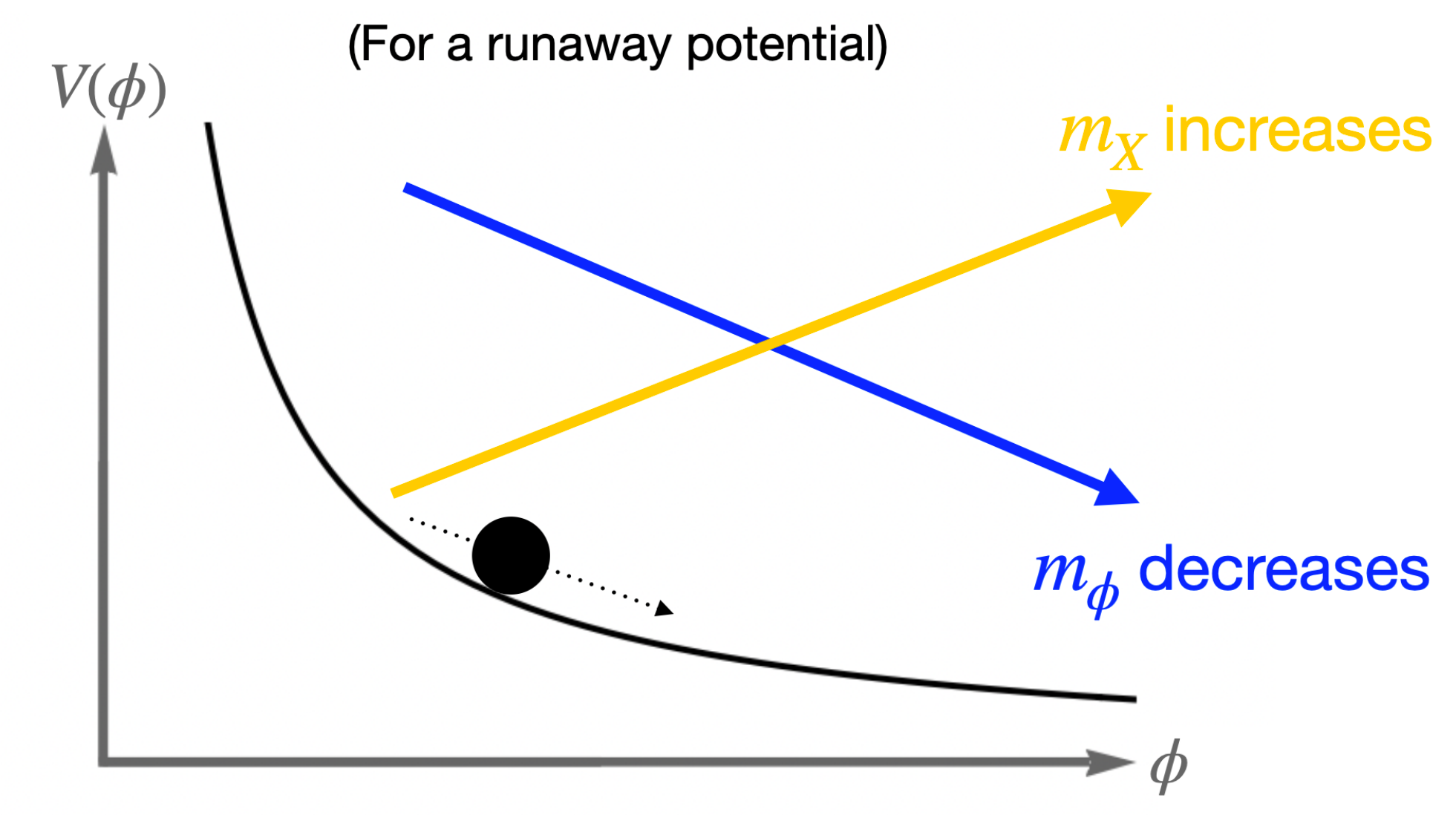}}
\caption{A schematic illustration of how the quintessence mass \(m_{\phi}\) and the dark gauge boson mass \(m_{X}\) evolve over cosmic time in the gauged quintessence model. The Ratra--Peebles potential drives \(\phi\) toward larger field values, causing \(m_{X}\) to increase while \(m_{\phi}\) decreases.}
\label{fig:mass_evolution}
\end{figure}

As \(\phi\) slowly rolls down the runaway potential \(V_0(\phi)\), the field value \(\phi(t)\) generally increases with cosmic time. Consequently, \(m_X(t)\) \emph{increases}, while \(m_{\phi}(t)\) \emph{decreases}. This interplay is illustrated schematically in Fig.~\ref{fig:mass_evolution}. To understand the energy flow between \(\phi\) and \(X\), one can derive the corresponding Boltzmann equations:
\begin{equation}
\begin{split}
\dot{\rho}_{\phi} + 3H\left(\rho_{\phi} + p_{\phi}\right) &= -\frac{\dot{m}_X}{m_X}\left(\rho_{X} - 3p_{X}\right), 
\\
\dot{\rho}_{X} + 3H\left(\rho_{X} + p_{X}\right)
&= \frac{\dot{m}_X}{m_X}\,\left(\rho_{X} - 3p_{X}\right),
\end{split}
\label{eq:finalboltz}
\end{equation}
where $\rho_\phi$ and $p_\phi$ are defined in Eq.~\eqref{eq:rho_phi & p_phi}, and the terms on the right-hand side originate from the gauge potential.
The quantities \(\rho_X\) and \(p_X\) are defined by \(\rho_X = \rho_{\phi+X} - \rho_{\phi}\) and \(p_X = p_{\phi+X} - p_{\phi}\), where \(\rho_{\phi+X}\) and \(p_{\phi+X}\) represent the total contributions from both \(\phi\) and \(X\). Notably, the right-hand sides of Eq.~\eqref{eq:finalboltz} feature equal and opposite terms proportional to \(\dot{m}_X/m_X\), indicating an energy exchange between the quintessence field and the dark gauge boson. If \(m_X\) \emph{increases} with time, energy flows from \(\phi\) into \(X\). Conversely, if \(m_X\) were to decrease, the flow would reverse. Due to the energy transfer between $X$ and $\phi$, the equation of state for quintessence in eq.~\eqref{eq:eos_quint} is not suitable for describing the evolution of the $\phi$ energy density. Instead, it is proper to consider the contribution from the energy transfer from $X$ as
\begin{equation}
    w_\phi = \frac{p_\phi}{\rho_\phi} + \frac{\dot{m}_X}{3H\rho_\phi m_X}(\rho_X-3p_X).
\end{equation}

\subsection{Combined Potential and Phenomenological Implications}
\label{subsec:GQ-potential}

\begin{figure}[tb]
\centerline{\includegraphics[width=0.8\linewidth]{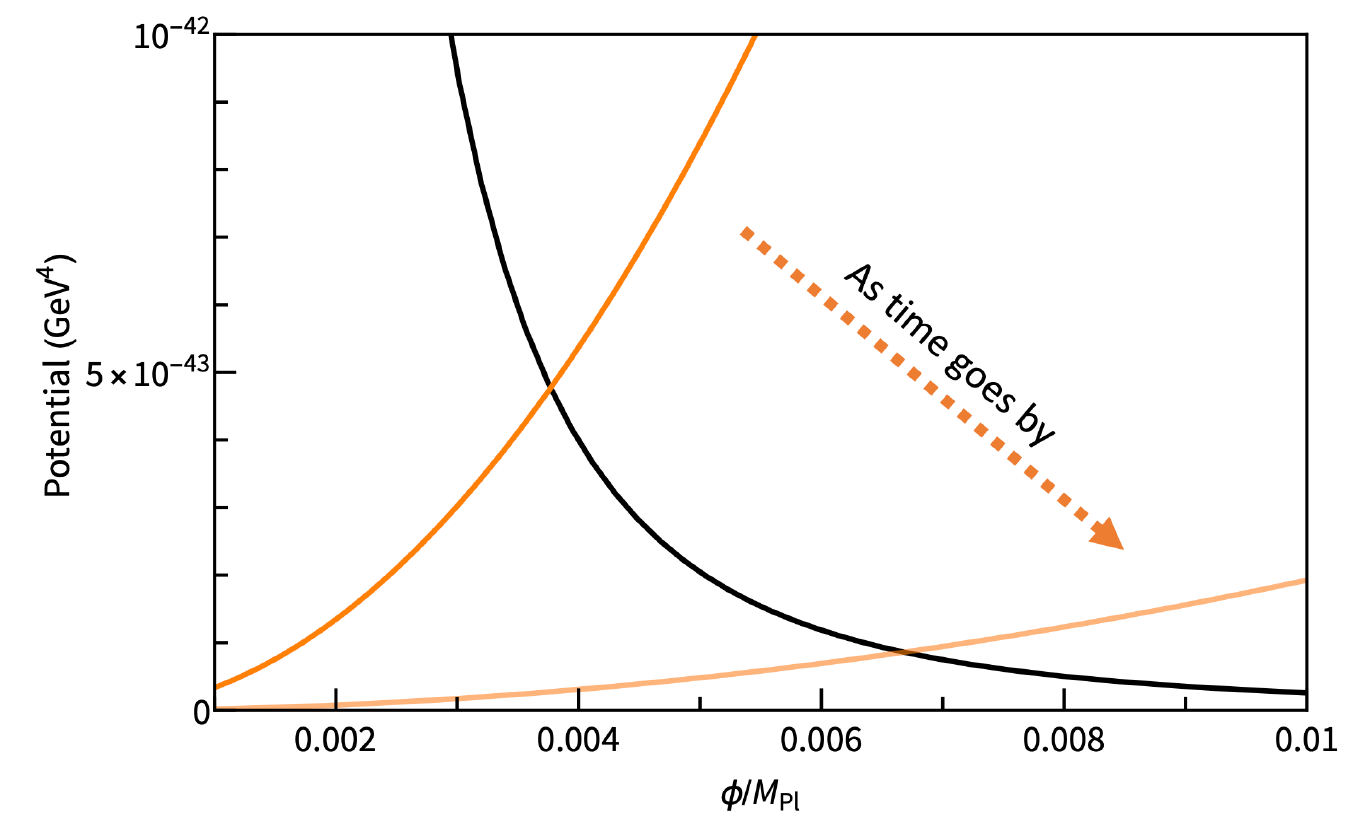}}
\caption{Schematic representation of the potential experienced by \(\phi\). The runaway Ratra--Peebles potential, \(V_{0}(\phi)\), is modified by the gauge potential, \(V_{\text{gauge}}(\phi)\).\cite{Kaneta:2022kjj} Their sum may admit distinct cosmological solutions compared to standard (ungauged) quintessence.}
\label{fig:V_gauge}
\end{figure}

Figure~\ref{fig:V_gauge} provides a qualitative sketch of how the \emph{gauge potential} \(V_{\text{gauge}}(\phi)\) combines with the original Ratra--Peebles potential \(V_0(\phi)\). The net effect is a competition between the ``repulsive'' runaway tendency in \(V_0(\phi)\) and the ``attractive'' mass term generated by \(V_{\text{gauge}}(\phi)\). Their sum defines the overall potential experienced by \(\phi\) and leads to rich cosmological phenomena, potentially including novel signatures that distinguish gauged quintessence from its standard, ungauged counterpart. In subsequent sections, we investigate how this interplay can influence the evolution of the universe, the production of dark gauge bosons, and potential observational consequences.

\section{Quantum Corrections and Constraints}
\label{sec:quantum_corrections}
In general, once a classical scalar field potential is established, quantum effects typically modify it. Consequently, the effective potential, incorporating all quantum corrections, can differ significantly from its classical counterpart. The Ratra-Peebles potential, in particular, is non-renormalizable and thus may be considered an effective potential. However, as discussed in Ref.~\citenum{Doran:2002bc}, this viewpoint could have subtle fine-tuning issues when quintessence fields interact with external fields. Specifically, particular relationships among these fields are required to precisely cancel large corrections arising from external interactions. In this section, we examine possible constraints on gauge couplings and the mass of the dark gauge boson by treating the Ratra-Peebles potential strictly as a classical potential.

\begin{figure}[tb]
\centering
\includegraphics[width=0.8\linewidth]{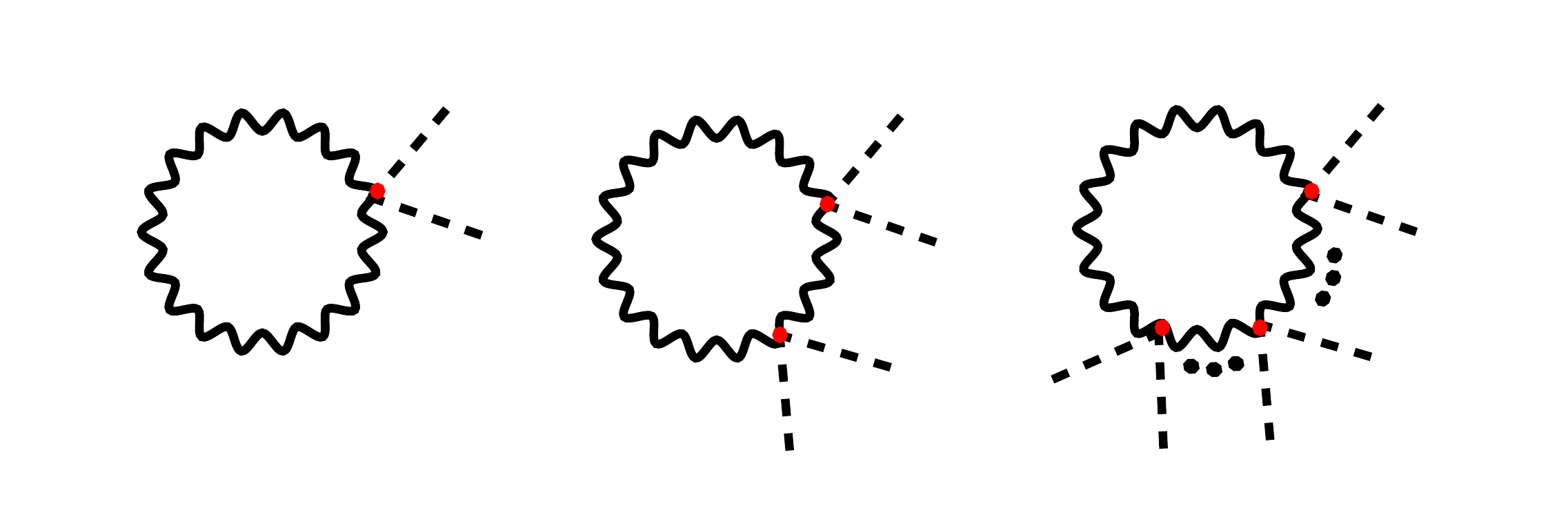}
\caption{Representative one-loop diagrams contributing to the effective potential from dark gauge boson loops.\cite{Kaneta:2022kjj}}
\label{fig:Feynman}
\end{figure}

We consider quantum corrections using the one-loop Coleman–Weinberg effective potential\cite{Coleman:1973jx,Jackiw:1974cv}. For concreteness, we incorporate both the scalar loop correction and the additional corrections induced by the dark gauge boson \(X\). The one-loop corrected potential is given by
\begin{align}
    V_{\text{eff}}(\phi) 
    =& 
    V_0(\phi) 
    + \frac{1}{2}g_X^2\phi^2X_\mu X^\mu 
    + \frac{\Lambda^2}{32\pi^2}V_0''(\phi)  \notag\\
    &
    +\frac{\left[V_0''(\phi)\right]^2}{64\pi^2} \left(\ln\left[\frac{V_0''(\phi)}{\Lambda^2}\right] - \frac{3}{2}\right)
    + \frac{3\left(m_{X}^2|_0\right)^2}{64\pi^2}
\left(\ln\left[\frac{m_{X}^2|_0}{\Lambda^2}\right] - \frac{5}{6}\right),
    \label{eq:QV}
\end{align}
where primes denote derivatives with respect to \(\phi\), and \(\Lambda\) is the renormalization scale. The first two terms correspond to the classical potential plus the gauge potential. The third and fourth terms are the one-loop scalar corrections, while the last term arises from the dark gauge boson loop (see Fig.~\ref{fig:Feynman}). Quadratic divergences from the gauge boson loop are absorbed into the scalar loop’s counterterm under a suitable fine-tuning scheme.

\subsection{Effective Quintessence Mass}
\label{subsec:eff_mass}

The physical (one-loop corrected) quintessence mass-squared can be obtained from the second derivative of \(V_{\text{eff}}\) with respect to \(\phi\):
\begin{align}
    m_{\phi}^2 
    =& 
    V_0''(\phi) 
    + g_X^2 \, X_\mu X^\mu
    + \frac{\Lambda^2}{32\pi^2}V_0''''(\phi)
    \notag\\
    &\quad +\frac{V_0''(\phi)V_0''''(\phi)}{32\pi^2}
    \left(\ln\left[\frac{V_0''(\phi)}{\Lambda^2}\right] - 1\right)
    +\frac{9g_X^2m_X^2|_0}{16\pi^2}
    \left(\ln\!\left[\frac{m_X^2|_0}{\Lambda^2}\right] + \frac{1}{3}\right).
    \label{eq:effphimass}
\end{align}
The first two terms represent the classical contribution and the tree-level gauge mass term, while the remaining terms are loop-level corrections. Note that these corrections must remain sufficiently small to ensure the slow-roll condition needed for quintessence.

\subsection{Constraints from Dark Energy Conditions}
\label{subsec:constraints}

In order for \(\phi\) to act as the dark energy field, both the potential \(V_{\text{eff}}(\phi)\) and the effective mass \(m_{\phi}^2\) must satisfy the requirements at the present epoch as discussed in Sec.~\ref{subsec:quintessence-mass}:
\begin{enumerate}
    \item \(\rho_{\phi} \approx V_{\text{eff}}(\phi_0) \simeq 3 \times 10^{-47}\,\mathrm{GeV}^4\), matching the observed dark energy density.
    \item \(m_{\phi} \lesssim H_0 \sim 10^{-42}\,\mathrm{GeV}\), ensuring the field remains slow-rolling on cosmological timescales.
\end{enumerate}
To estimate the parameter space that satisfies these conditions, we employ a conservative approach: we require each term in \(V_{\text{eff}}\) and \(m_{\phi}^2\) to individually respect these bounds rather than allowing cancellations among loop corrections. Although such cancellations could, in principle, occur via fine-tuning, they are expected to be unstable over cosmic timescales.

\subsection{Implications for Gauge Coupling and Dark Gauge Boson Mass}
\label{subsec:gauge_constraint}

\begin{figure}[tb]
\centering
\includegraphics[width=0.8\linewidth]{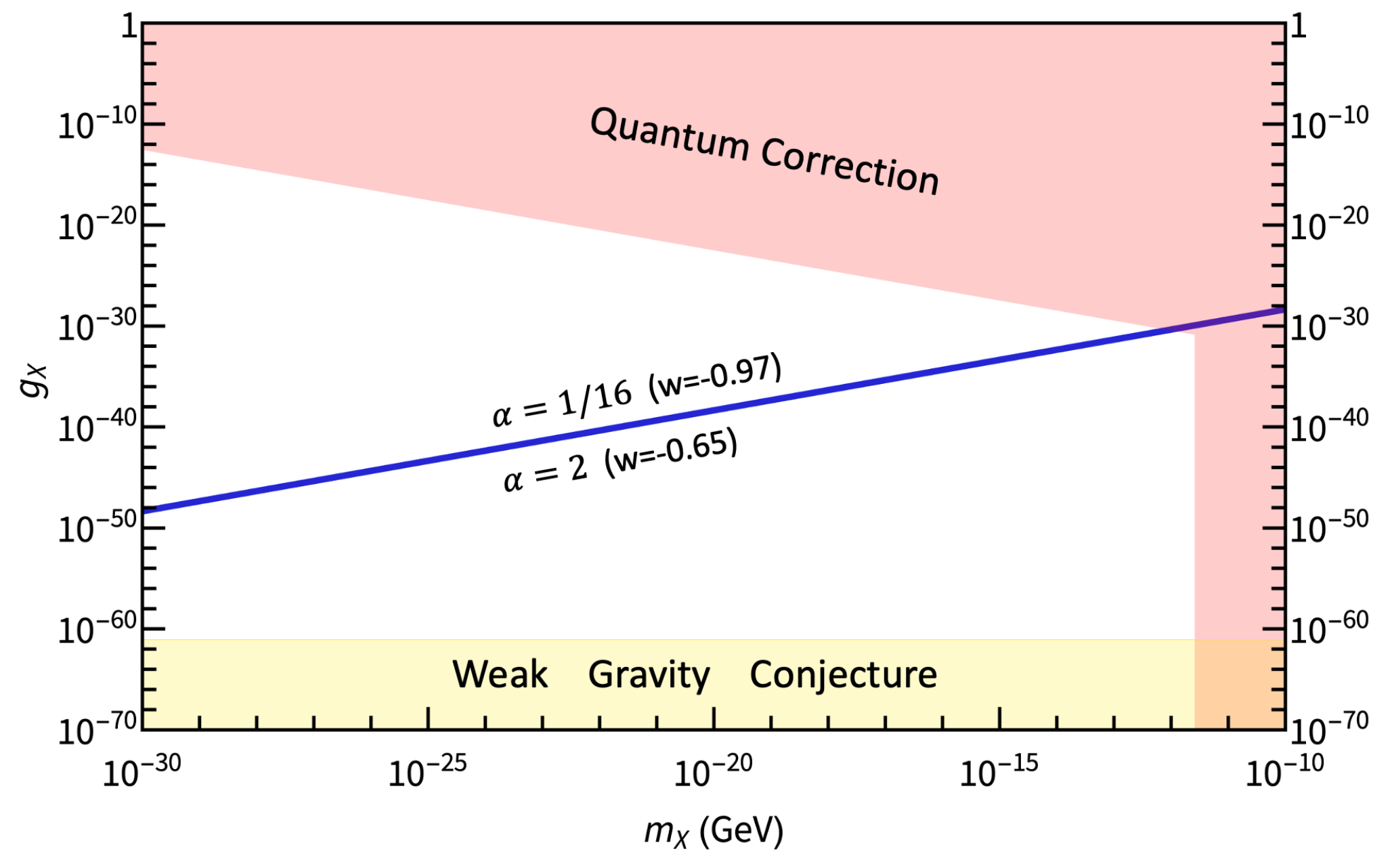}
\caption{Constraints on the dark gauge coupling \(g_X\) and the dark gauge boson mass \(m_X\).\cite{Kaneta:2022kjj} The red region is excluded by quantum-correction arguments discussed in Section~\ref{subsec:constraints}, while the blank region is generally allowed. The narrow blue band represents the case of the Ratra--Peebles inverse-power-law potential with tracking behavior.}
\label{fig:loop_constraint}
\end{figure}

Figure~\ref{fig:loop_constraint} illustrates the resulting constraints on the dark gauge coupling \(g_X\) and the dark gauge boson mass \(m_X\). The red-shaded region indicates the exclusion derived from requiring the loop corrections to remain below the dark energy scale and mass constraints at the current epoch. The unshaded (white) region is broadly allowed without reference to a specific choice of \(V_0(\phi)\). For the particular case of the Ratra--Peebles potential with tracking behavior, the viable parameter space corresponds to the narrow blue band. 

In summary, quantum corrections may impose meaningful restrictions on gauged quintessence models, particularly when requiring that the effective potential and mass remain compatible with an ultra-light field driving cosmic acceleration. These bounds can be satisfied in wide regions of parameter space, with or without specific assumptions on the form of the classical potential \(V_0(\phi)\).

\section{Misalignment Mechanism for Vector Boson Production}
\label{sec:misalignment}
We now consider the production of the dark gauge boson \(X\) in the early universe via the \emph{misalignment mechanism}. Although multiple processes could generate \(X\), we focus here on the scenario in which a coherent homogeneous field of \(X\) is established through inflation and subsequently evolves as the universe expands.

\subsection{Review: Misalignment for a Coherent Scalar Field}
\label{subsec:scalar_misalignment}

It is instructive to begin with the well-known misalignment mechanism for a light scalar field \(\varphi\)\cite{Preskill:1982cy,Abbott:1982af,Dine:1982ah,Arias:2012az}. In this case, one typically studies the equation of motion in the expanding universe,
\begin{equation}
    \ddot{\varphi} 
    + 3H \dot{\varphi} 
    + m_{\varphi}^2\,\varphi 
    \;=\; 0,
    \qquad
    \rho_{\varphi} 
    \;=\; 
    \frac12\,\dot{\varphi}^2 
    + \frac12\,m_{\varphi}^2\,\varphi^2,
    \label{eq:EOM for varphi}
\end{equation}
where \(m_{\varphi}\) is the (constant) mass of the scalar. During inflation, spatial inhomogeneities in \(\varphi\) are stretched to super-Hubble scales, resulting in a nearly uniform field value within our observable patch, thus justifying Eq.~\eqref{eq:EOM for varphi}. If \(H \gg m_{\varphi}\) at early times, Hubble friction freezes \(\varphi\) away from the minimum of its potential, preserving a significant energy density. Once the expansion rate drops below \(m_{\varphi}\), the field begins coherent oscillations around the potential minimum. Because these oscillations have pressure \(p_{\varphi} \simeq 0\), the energy density \(\rho_{\varphi}\) redshifts like nonrelativistic matter \(\propto a^{-3}\). This mechanism is a key ingredient in the cosmology of ultra-light scalars such as the QCD axion.

\begin{figure}[tb]
\centering
\includegraphics[width=0.8\linewidth]{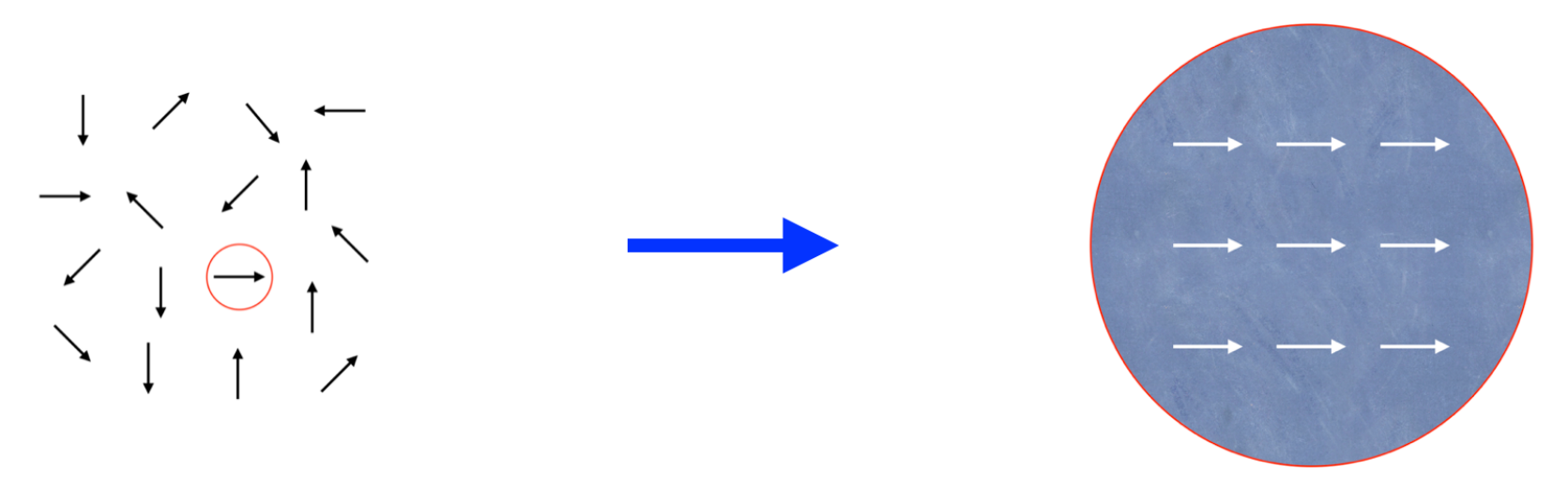}
\caption{A schematic depiction of a coherent vector field mode. Before inflation, the vector fields have random orientations, but inflation aligns them in a single direction across the observable universe. Consequently, one can treat \(X_\mu\) as a time-dependent yet spatially uniform field.}
\label{fig:vectorField}
\end{figure}

\subsection{Extension to a Coherent Vector Field}
\label{subsec:vector_misalignment}

We now turn to the analogous scenario for a massive spin-1 field \(X\)\cite{Nelson:2011sf}. Before inflation, one might imagine that the vector field values are oriented randomly in different patches of the universe. After sufficient inflation, however, the observable universe is dominated by a single direction of \(X\) (see Fig.~\ref{fig:vectorField}). We focus on the zero-momentum mode, treating
\begin{equation}
    X_\mu(t,\vec{x}) = X_\mu(t),
\end{equation}
so that the field is spatially homogeneous (coherent). In such a configuration, one can have only \(X_0 = 0\). A single spatial component (say, $z$-direction) then obeys an equation of motion analogous to the scalar case:
\begin{align}
    \ddot{X} & + H\dot{X} + m_X^2X = 0, \\
    \rho_X & = \frac{1}{2a^2}\left(\dot{X}^2 + m_X^2X^2\right).
\end{align}
Here \(m_X\) is the mass of the vector, taken to be constant in typical scenarios. Unlike a scalar case, \(\rho_X\) suffers an additional suppression by the factor \(1/a^2\), which rapidly diminishes any primordial amplitude through inflation (up to \(\sim 60\) \(e\)-folds or more). As a result, the naive misalignment mechanism for a strictly constant-mass vector typically fails to generate a significant relic abundance.

\subsection{Misalignment with a Mass-Varying Vector}
\label{subsec:mass_varying_vector}

In the gauged quintessence scenario, however, the dark gauge boson mass is \emph{not} fixed. Instead, it is controlled by the value of the quintessence field,
\begin{equation}
    m_X^2(t) = g_X^2\,\phi^2(t),
\end{equation}
which can vary by many orders of magnitude throughout cosmic history. Consequently, even if \(\rho_X\) becomes initially suppressed by inflation, the increasing mass \(m_X\) can compensate the dilution during the inflation, allowing a non-negligible \(\rho_X\) to develop via the misalignment mechanism.

\begin{figure}[tb]
\centering
\includegraphics[width=0.8\linewidth]{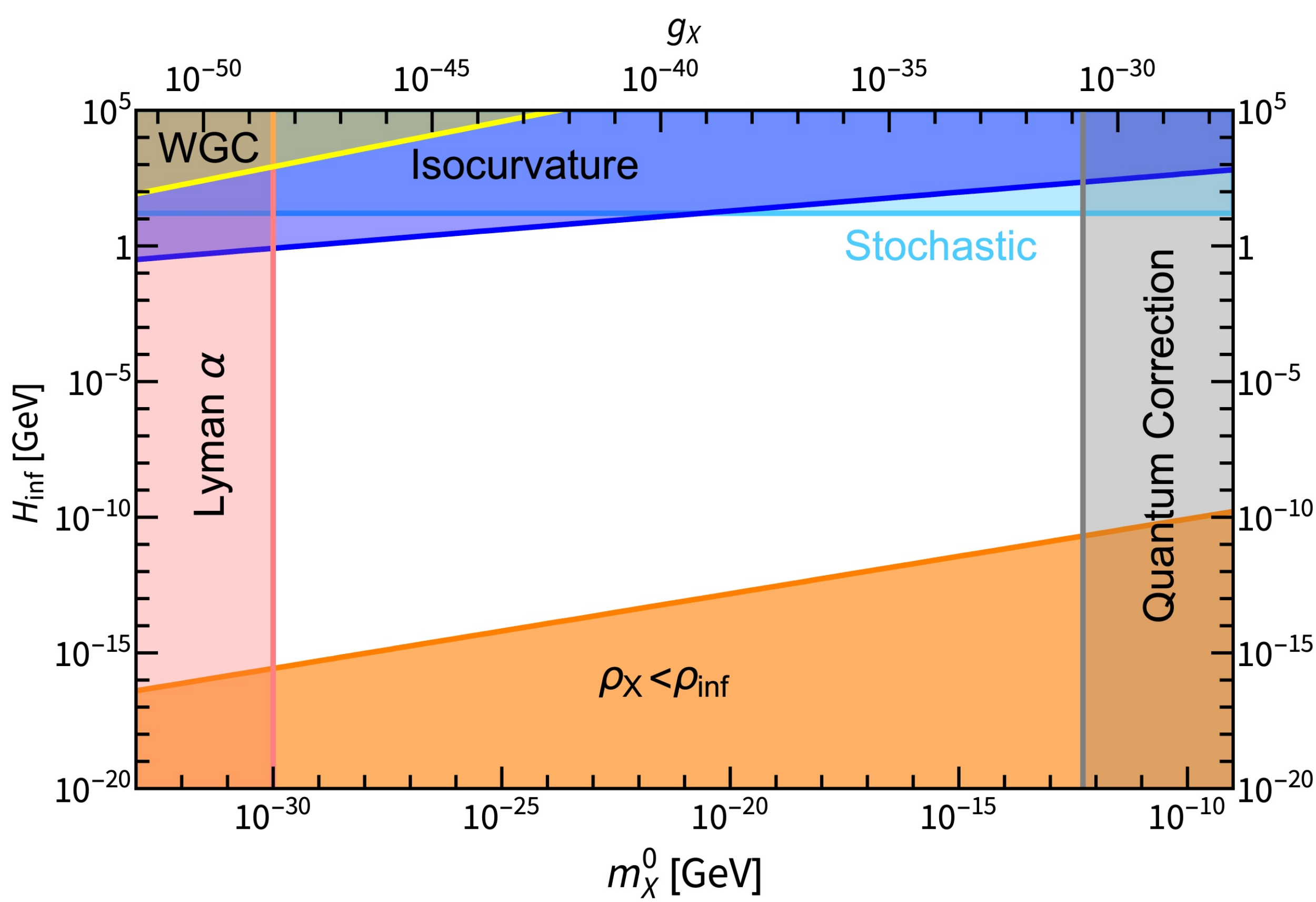}
\caption{An illustration of the parameter space in \(\bigl(m_X^0,\,H_{\text{inf}}\bigr)\) where a coherent vector boson can be produced via misalignment in the gauged quintessence model.\cite{Kaneta:2023lki} Here \(H_{\text{inf}}\) is related to the reheating temperature or the energy scale during inflation. Constraints from quantum corrections, relic abundance, and other cosmological observations limit the colored regions; the unshaded regions remain viable. The viable parameter space depends on the fraction of \(X\) in the present-day relic density. Here, we illustrate the scenario in which \(X\) density is comparable to the relic density of cold dark matter.}
\label{fig:misalignment_paramspace}
\end{figure}

Figure~\ref{fig:misalignment_paramspace} shows a representative parameter space for \(m_X^0\) (the present-day dark gauge boson mass) versus \(H_{\mathrm{inf}}\) (the inflationary Hubble scale). Various constraints—such as those from quantum corrections (Section~\ref{sec:quantum_corrections}), isocurvature bounds, and relic abundance considerations—restrict portions of this space.\cite{Kaneta:2023lki} Nevertheless, sizeable allowed regions remain, where the mass-varying vector boson can be produced via the misalignment mechanism and potentially constitute a portion of the dark sector.

In summary, while the misalignment mechanism for a constant-mass vector is typically negligible, the \emph{gauged quintessence} framework circumvents this suppression. The key ingredient is the time-dependence of \(m_X(t)\) driven by the rolling quintessence field \(\phi\). This opens an interesting window for dark gauge bosons as potential constituents of dark matter or other cosmologically relevant relics, which we discuss further in subsequent sections.

\section{Evolution of the Universe}
\label{sec:evolution}
We now turn to the cosmological evolution in gauged quintessence, focusing on how the coupled scalar field \(\phi\) and dark gauge boson \(X\) evolve through different epochs. 
The coupled equations of motion for \(\phi\) and \(X\) (in a spatially homogeneous approximation) are:
\begin{equation}
\begin{split}
    &\ddot{\phi} + 3H\dot{\phi} + \frac{\partial V_0}{\partial \phi} 
    + g_X^2\left(X_\mu X^\mu\right)\phi 
    = 0,\\
    & \ddot{X} + H\dot{X}+g_X^2\phi^2X = 0 .
\end{split}
\label{eq:eom}
\end{equation}
Here \(V_0(\phi)\) is the original Ratra--Peebles potential, while the last terms in both equations capture the gauge potential term. As \(\phi\) evolves, it modifies \(m_X^2 = g_X^2\,\phi^2\), thus affecting the vector field evolution as well.  Due to the coupled nature of \(\phi\) and \(X\), (see eq.~\eqref{eq:finalboltz}), numerical evaluation is necessary to determine their evolution. 

\begin{figure}[tb]
\centering
\includegraphics[width=0.8\linewidth]{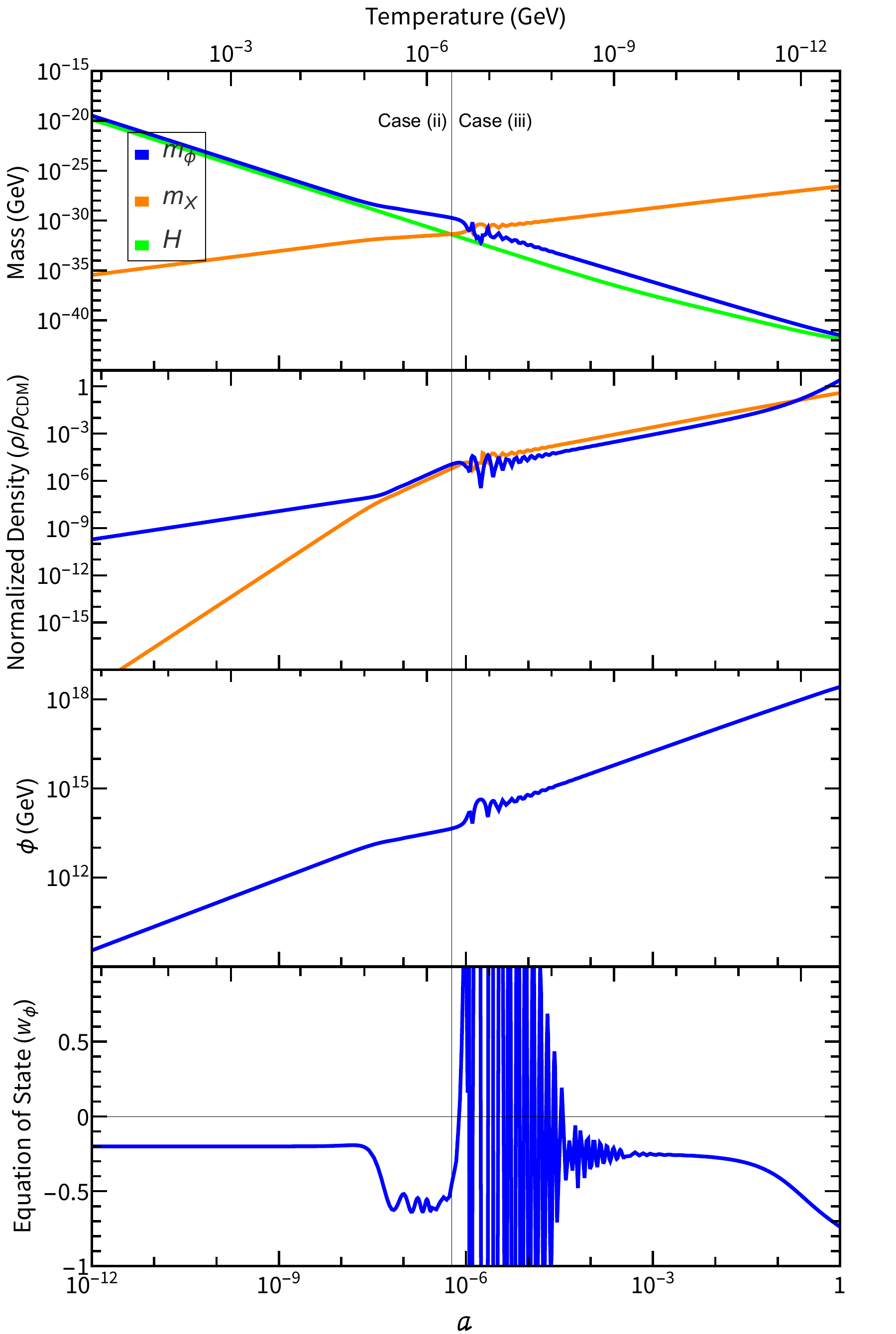}
\caption{An illustration of how \(\phi\) and \(X\) evolve through various stages of cosmic expansion.\cite{Kaneta:2022kjj} 
The vertical lines mark the transitions in the hierarchy among \(m_{X}\), and the Hubble parameter \(H\).
The parameters shared by all curves are \(g^{}_X = 10^{-45}\), \(\alpha=1\), \(M=2.2\times 10^{-6}\,\text{GeV}\), and \(\Lambda=M^{}_{\text{Pl}}\). Furthermore, \(\rho_X/\rho_{\text{CDM}}\bigl|_i = 7 \times 10^{-20}\) at \(a=10^{-12}\) (the initial fraction of the dark gauge boson energy density to the CDM energy density). The current Hubble parameter is taken as \(H_0=67.4\,\mathrm{km\,s^{-1}\,Mpc^{-1}}.\) }
\label{fig:evolutions}
\end{figure}

\subsection{Evolutionary Stages}
\label{subsec:evolution_stages}

In the evaluation of eq.~\eqref{eq:eom}, hierarchy among \(\{m_{\phi},\,m_{X},\,H\}\) changes the evolution of $\phi$ and $X$ significantly. We identify three distinct regimes:

\begin{enumerate}
    \item \textbf{Case (i):} \(H \gg m_{\phi},\,m_{X}\). 
    \\
    Both \(\phi\) and \(X\) are effectively frozen by Hubble friction, remaining nearly constant in time. In this case, the evolution of both fields are trivial. 

    \item \textbf{Case (ii):} \(m_{\phi} \gtrsim H \gg m_{X}\).  
    \\
    The quintessence field \(\phi\) rolls down its potential, while the $X$ field is effectively frozen (\(m_{X} \ll H\)). Trapped regime, where $\phi$ follows the minimum of the potential (see Fig.~\ref{fig:V_gauge}), is achieved when the background $X$ density is large ($m_\phi \gg H$), or $\phi$ field follows the tracking solution when $m_\phi \sim H$. 

    \item \textbf{Case (iii):} \( m_\phi,\,m_{X} \gtrsim H\).  
    \\
    While the dynamics of $\phi$ field can be described using either a trapped or tracking solution, $X$ field undergoes rapid oscillations with an oscillation timescale of $1/m_X$. Therefore, the equation of motion for \(X\) is difficult to solve numerically; instead, its evolution is approximated by a WKB solution\cite{Kaneta:2022kjj},
    \begin{equation}
 X_i(t) \propto \text{Re} \left[\frac{\chi^{}_i}{\sqrt{a m_X^{}}} e^{i \int dt\,  m_X^{}} \right].
    \end{equation}
    
\end{enumerate}

\subsection{Illustrative Numerical Results}
\label{subsec:numerical_evolution}

We present the numerical results for various quantities associated with the $X$ and $\phi$ fields.
For simplicity, we assume that \(X\) constitutes only a fraction of the total dark matter, to avoid complications arising from a fully mass-varying dark matter scenario.

\paragraph{Mass Hierarchies.}
In the first panel of Fig.~\ref{fig:evolutions}, we show how \(H\), \(m_{\phi}\), and \(m_{X}\) evolve over many orders of magnitude in scale factor, where the evolution starts from \textbf{Case (ii)} to \textbf{Case (iii)}. The vertical lines mark the transitions between two stages.

\paragraph{Energy Densities, Field evolution, Equation of state}
The subsequent three panels show the normalized energy densities of \(\phi\) and \(X\), the field evolution, and the equation of state, respectively. In the later part of \textbf{Case (ii)}, \(\rho_X\) becomes comparable to $\rho_\phi$, allowing the gauge potential to affect the evolution of $\phi$. In this regime, $\phi$ may have mild oscillations along the trapped potential, as seen from the third and fourth panels of Fig.~\ref{fig:evolutions}. As the evolution transitions into \textbf{Case (iii)}, where $m_X \gtrsim H$, the $X$ field starts oscillation, producing strong oscillation patterns in the $\phi$ field. This leads to a violent energy transfer between $X$ and $\phi$, reflected in the oscillations of $w_\phi$. Notably, the value of $w_\phi$ can extend beyond the typical range of $[-1,1]$ due to the energy exchange between $\phi$ and $X$.

\subsection{Summary of Cosmological Evolution}
\label{subsec:evolution_summary}

These numerical examples highlight the key feature of gauged quintessence: both \(\phi\) and \(m_X\) vary significantly over cosmic time. The interplay between the runaway quintessence potential \(V_0(\phi)\) and the gauge potential $V_\text{gauge}$ gives rise to a rich cosmological history. In particular, the mass of the dark gauge boson evolves by many orders of magnitude, opening the door for unique phenomenological signatures, including a transient era where \(\rho_X\) may be sizeable. Nevertheless, by assuming \(X\) is a subdominant dark matter component, one can avoid many of the more stringent constraints that would arise if it were the primary dark matter candidate.

In the following sections, we discuss further implications of gauged quintessence, including the possibility of addressing the Hubble tension and exploring non-gravitational signals of a dynamically evolving dark gauge sector.

\section{Hubble Tension}
\label{sec:hubble_tension}
A persistent discrepancy in measurements of the Hubble constant \(H_0\) has emerged in recent years, dubbed the ``Hubble tension.'' Specifically, early-Universe probes (such as fits to the cosmic microwave background (CMB) data under \(\Lambda\)CDM) favor \(H_0 \approx 67\,\mathrm{km\,s^{-1}\,Mpc^{-1}}\)\cite{Planck:2018vyg}, whereas direct late-Universe measurements (e.g., using standard candles in the local universe) indicate \(H_0 \approx 73\,\mathrm{km\,s^{-1}\,Mpc^{-1}}\)\cite{Riess:2021jrx}.  The mismatch between these two primary methods suggests new physics beyond the standard \(\Lambda\)CDM model or unrecognized systematic errors in one (or both) measurements.

\subsection{BAO Constraints and Angular Sound Horizon}
\label{subsec:BAO_constraints}

Any new dark energy model that attempts to resolve the Hubble tension must be compatible with baryon acoustic oscillation (BAO) data. The BAO could be measured in the CMB or the distributions of astrophysical objects. Formally, this angle is determined by the ratio between:

\begin{enumerate}
    \item The \emph{sound horizon} \(r_s\), which depends on the early-Universe physics (before the last-scattering surface),
    \item The \emph{comoving distance} \(D(z)\) to the object, determined by late-Universe expansion.
\end{enumerate}

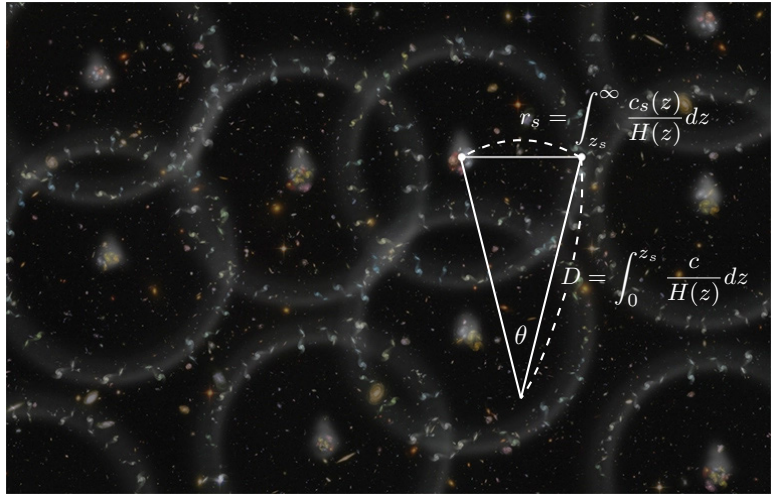
\begin{figure}[tb]
\centering
\begin{tikzpicture}
\node at (0,0) {\includegraphics[width=.8\linewidth]{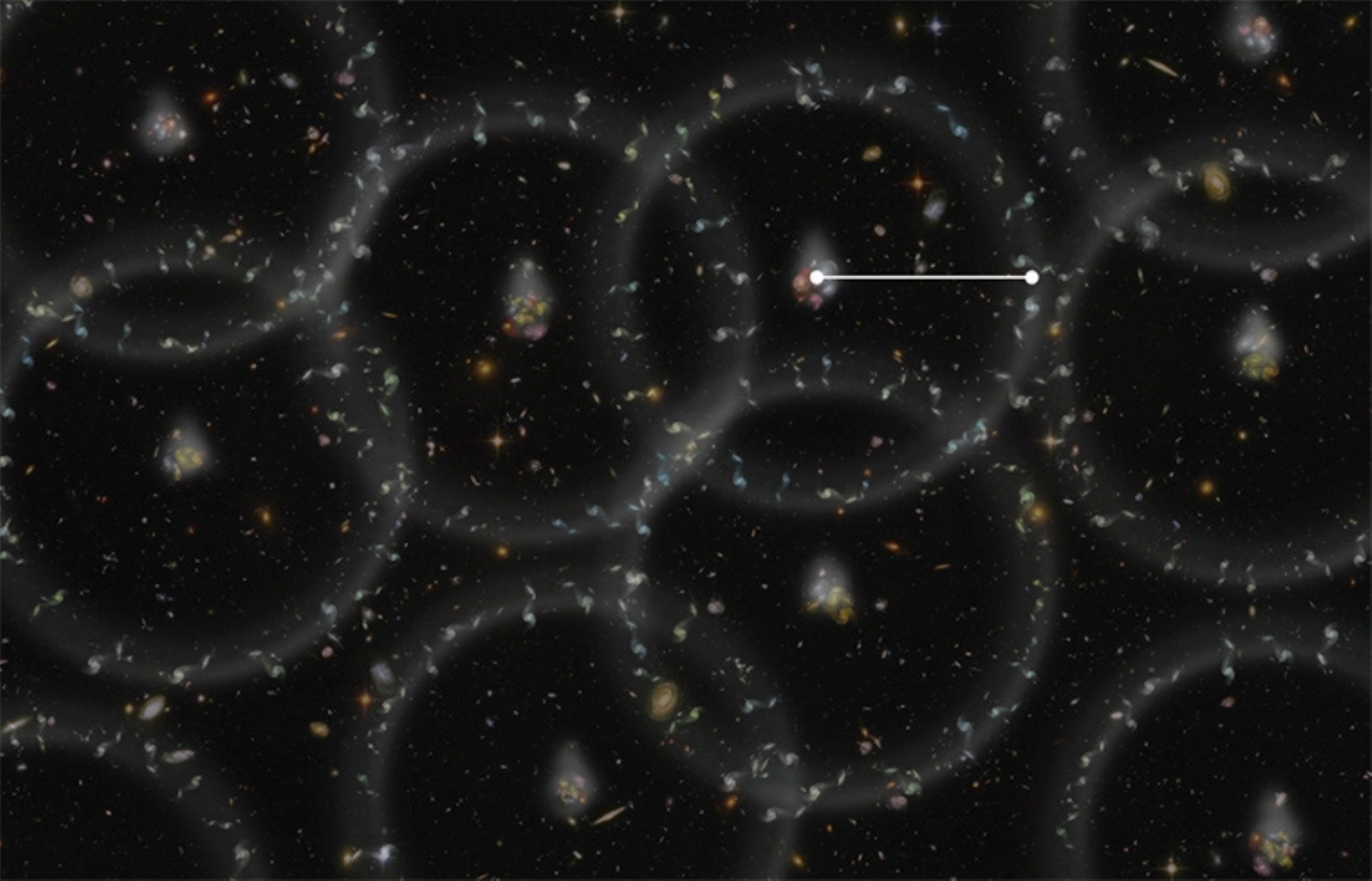}};
    \draw[thick, white, fill=white] (.96,1.2) -- (1.76,-2);
    \draw[thick, white, fill=white] (2.56,1.2) -- (1.76,-2);
    \draw[thick, white, dashed] (2.56,1.2) to[bend right] (.96,1.2);
    \draw[thick, white, dashed] (2.56,1.2) .. controls (2.6,0.6) and (2.55,0.3) .. (2.45,-.2);
    \draw[thick, white, dashed] (2.35,-.6) .. controls (2.28,-.8) and (2,-1.7) .. (1.76,-2);
    \node[above] at (1.76,-1.4) {\color{white} $\theta$};
    \node[above] at (2.9,1.2) {\small \color{white} $\displaystyle\ \   r_s =  \int_{z_s}^{\infty} \frac{c_s(z)}{H(z)} dz $};
    \node[right] at (2.16,-.4) {\small \color{white} $\displaystyle   D = \int_0^{z_s}  \frac{c}{H(z)}dz $};
\end{tikzpicture}
\caption{A schematic depiction of how the BAO scale \(\theta\) constrains new-physics scenarios. The BAO peak results from the interplay of the sound horizon \(r_s\) (early universe) and the comoving distance \(D(z_s)\) (late universe). The background image is taken from Ref.~\citenum{baoimg}.}
\label{fig:BAO}
\end{figure}

In simpler terms,
\begin{equation}
\begin{split}  
    r_s &= \int_{z_s}^{\infty} \frac{c_s(z)}{H(z)} \,dz, 
    \\
    D(z) &= \int_{0}^{z} \frac{c}{H(z)} \,dz 
    \;=\; \frac{r_s}{\theta},
\end{split}
\label{eq:soundHorizon}
\end{equation}
where \(c_s(z)\) is the sound speed prior to recombination, and \(\theta\) is the observed angular scale of the BAO. (See Fig.~\ref{fig:BAO} for an illustration.)
Dark energy became dominant only in the late universe. Unless new physics is introduced in the early universe to modify the sound horizon \(r_s\) (early-universe physics), the comoving distance to the last scattering \(D(z_s)\) (late-universe physics) should remain unchanged under a new dark energy model.

\subsection{Why \(w < -1\) at Late Times?}
\label{subsec:w_less_than_-1}

A critical insight from BAO considerations is that to reconcile a larger value of \(H_0\) with an unchanged \(D(z)\), one typically requires a \emph{lower} Hubble expansion rate \(H(z)\) in the more recent past.\cite{DiValentino:2016hlg,DiValentino:2017iww,Joudaki:2017zhq,Heisenberg:2022lob,Heisenberg:2022gqk,Lee:2022cyh} $D(z_s)$ is the integration of $c/H(z)$ which becomes smaller at $z=0$ for preferring to larger Hubble constant. To compensate it for fixed $D(z_s)$, $H(z)$ should be smaller in the recent past. Mathematically, since
\[
H^2(z) \;=\; H_0^2\,\Bigl[\Omega_{\mathrm{DE}}(z) + \Omega_{\mathrm{m}}(z)\Bigr],
\]
and \(\rho \propto a^{-3(1+w)}\), it follows that a higher present-day \(H_0\) can be balanced by a period in which \(\rho_{\mathrm{DE}}\) evolves more slowly or even \emph{increases} with decreasing redshift. This generally points to an effective equation of state \(w_{\mathrm{eff}} < -1\) for part of the late-time evolution. Unfortunately, uncoupled quintessence has \(w \geq -1\), making the tension \emph{worse} compared to \(\Lambda\)CDM.\cite{Banerjee:2020xcn,Lee:2022cyh}

\subsection{Effective Equation of State in Gauged Quintessence}
\label{subsec:w_eff_gauged}

Several authors have explored \emph{interacting} dark energy scenarios in which \(w_{\mathrm{eff}}\) can dip below \(-1\), potentially alleviating the Hubble tension. In our gauged quintessence framework, the dynamical mass of the dark gauge boson \(m_X^2(t) = g_X^2\,\phi^2(t)\) induces an energy flow between \(\phi\) and \(X\). As a result, the effective dark energy density \(\rho_{\widetilde{\mathrm{DE}}}\) can behave differently from standard quintessence.

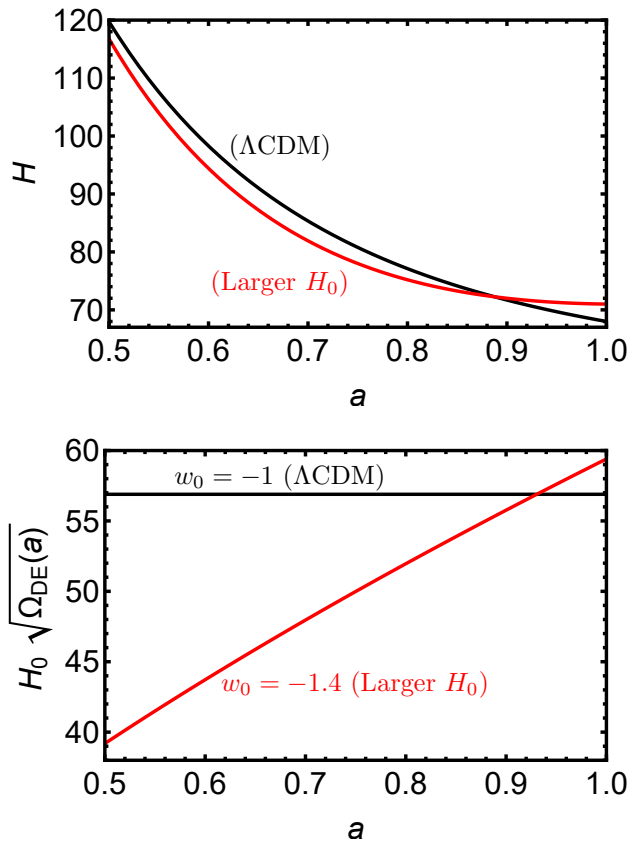
\begin{figure}[tb]
\centering
\begin{tikzpicture}
    \node at (0,0) {\includegraphics[width=.65\linewidth]{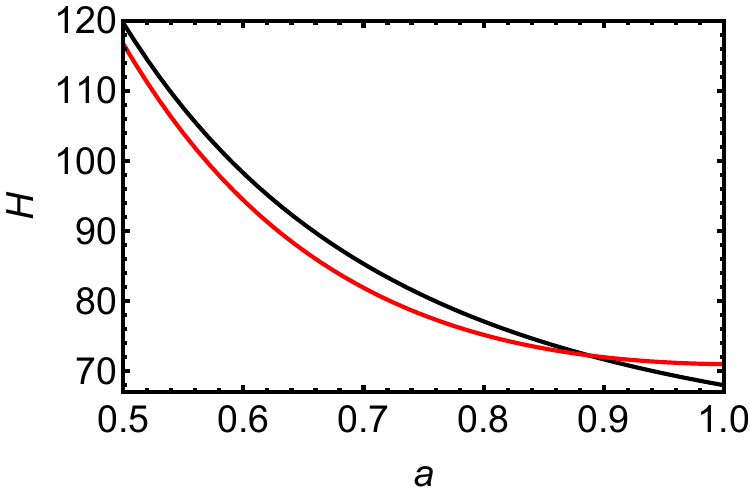}}; 
    \node at (-.5,0.8) {($\Lambda$CDM)};
    \node at (-.5,-1) {\color{red} (Larger $H_0$)};
\end{tikzpicture}
\begin{tikzpicture}
    \node at (0,0) {\includegraphics[width=.65\linewidth]{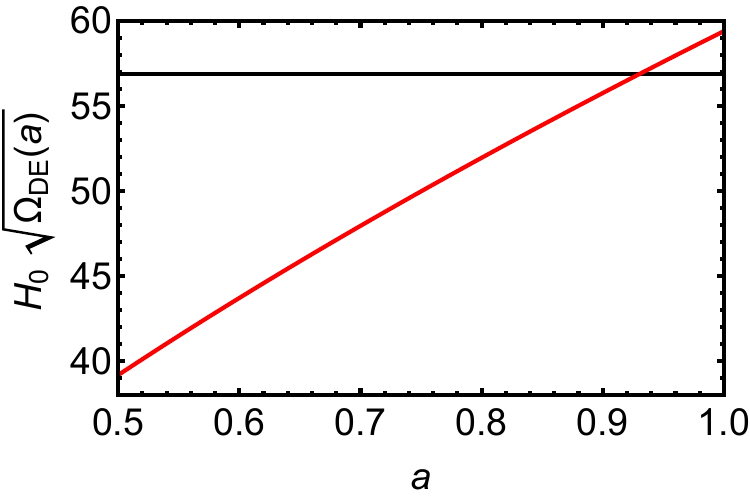}}; 
    \node at (-.5,2.15) {$w_0=-1$ ($\Lambda$CDM)};
    \node at (0.5,-0.6) {\color{red} $w_0=-1.4$ (Larger $H_0$)};
\end{tikzpicture}
\caption{Conceptual sketch of the expansion rate \(H(a)\) in \(\Lambda\)CDM (black curves) and in a hypothetical dark energy model that reduces \(H(a)\) at intermediate redshifts, then yields a higher \(H_0\). Such a scenario may mitigate the Hubble tension but must remain consistent with the BAO scale.}
\label{fig:Hubble}
\end{figure}

Figure~\ref{fig:Hubble} provides a schematic illustration of how \(H(z)\) might differ from \(\Lambda\)CDM (black curves) if an extra interaction shifts energy between \(\phi\) and other components. Meanwhile, Fig.~\ref{fig:weff} shows the effective equation of state parameter \(w_{\mathrm{eff}}(\widetilde{\mathrm{DE}})\) for gauged quintessence under various assumptions about the fraction of \(X\) in the total dark matter. A convenient expression is
\begin{equation}
    w_{\text{eff}}\left(\widetilde{\mathrm{DE}}\right)
    =
    -1 
    + \frac{1}{\rho_{\widetilde{\mathrm{DE}}}}
    \left[(1 + w_0)\rho_{\phi} 
        + \left(\frac{m_X}{m_X^0}-1\right)
        \frac{\rho_X^{0}}{a^3}\right],
    \label{eq:effwDEeff}
\end{equation}
where \(w_0 = -1\) in \(\Lambda\)CDM, \(m_X^0\) and \(\rho_X^0\) are the present-day mass and density of \(X\), and \(\rho_{\widetilde{\mathrm{DE}}}\) includes both \(\phi\) and the portion of \(X\) that effectively contributes to late-time acceleration. Because \(m_X\) was smaller in the past (\(m_X < m_X^0\)), the overall correction in Eq.~\eqref{eq:effwDEeff} can be negative, potentially driving \(w_{\mathrm{eff}}\) below \(-1\).

\subsection{Numerical Indications and Outlook}
\label{subsec:hubble_outlook}

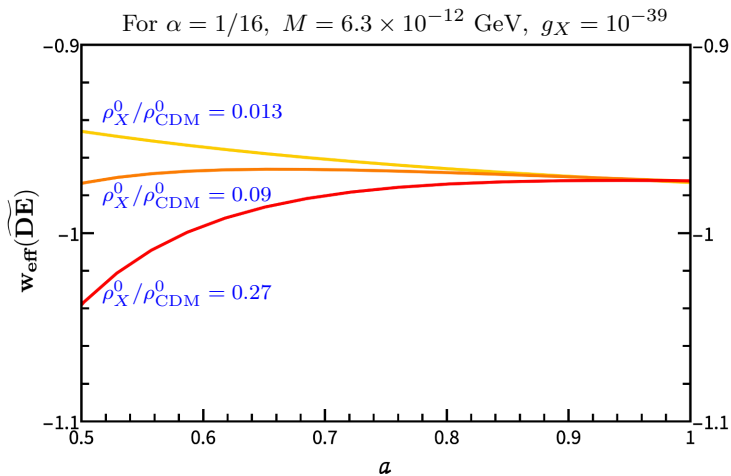
\begin{figure}[tb]
\centering
\begin{tikzpicture}
\node at (0,0) {\includegraphics[width=0.8\linewidth]{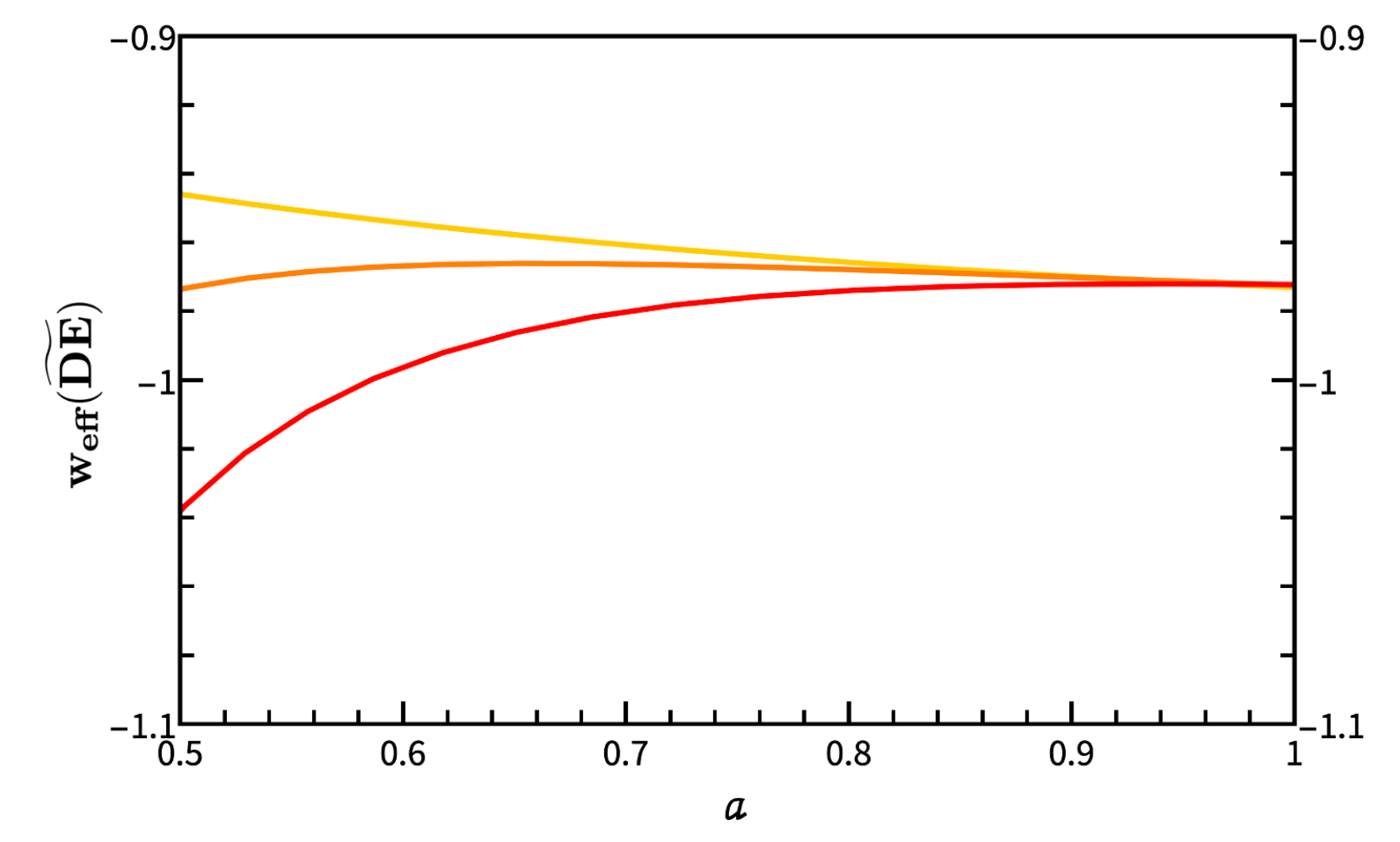}};
\node at (0.4,3.1) {\small For $\alpha=1/16, \ M = 6.3 \times 10^{-12} \text{ GeV},\ g_X=10^{-39} $};
\node at (-2.3,1.9) {\footnotesize \color{blue} $\rho_X^0/\rho_\text{CDM}^0 = 0.013$};
\node at (-2.3,0.8) {\footnotesize \color{blue} $\rho_X^0/\rho_\text{CDM}^0 = 0.09\color{white}0$};
\node at (-2.3,-0.5) {\footnotesize\color{blue} $\rho_X^0/\rho_\text{CDM}^0 = 0.27\color{white}0$};
\end{tikzpicture}
\caption{The effective equation of state \(w_{\mathrm{eff}}\bigl(\widetilde{\mathrm{DE}}\bigr)\) for gauged quintessence.\cite{Kaneta:2022kjj} Different colored lines denote varying fractions of \(X\) in the total dark matter content. Values dipping below \(-1\) in recent cosmological time (\(z \lesssim 1\)) may help ease the Hubble tension.}
\label{fig:weff}
\end{figure}

In Fig.~\ref{fig:weff}, we show that for a sufficiently large fraction of \(X\) within the dark matter component, \(w_{\mathrm{eff}}\bigl(\widetilde{\mathrm{DE}}\bigr)\) can indeed dip below \(-1\). This behavior is absent in uncoupled quintessence and may help reconcile the discrepancy between early and late measurements of \(H_0\). A detailed resolution of the Hubble tension, however, would require a full numerical fit to CMB data and other large-scale structure constraints, which we reserve for future work.

In summary, gauged quintessence could potentially serve as a concrete example of the interacting dark energy model that addresses the Hubble tension. By allowing energy transfer between \(\phi\) and a mass-varying vector boson \(X\), the effective equation of state for dark energy can be driven below \(-1\) for a period, lowering \(H(z)\) at intermediate redshifts while retaining a larger final value \(H_0\). Verifying this conjecture demands a systematic comparison with precision cosmological data, including BAO and CMB measurements.

\section{Non-Gravitational Signals}
\label{sec:signals}
Thus far, our focus has been on the gravitational and cosmological aspects of gauged quintessence. However, if the dark energy sector involves a gauged \(U(1)\), it is natural to ask whether non-gravitational signals might arise via a \emph{portal} interaction. Portals between the dark and visible sectors are commonly studied in dark matter phenomenology;\cite{Essig:2013lka} here, we explore the implications of a \emph{vector portal} for gauged quintessence. To facilitate these signals, we focus on a somewhat altered scenario in which:
1) \(X\) is not the dark matter component, and 
2) \(V_0(\phi)\) is treated as the \emph{quantum}-corrected potential.

\begin{figure}[tb]
\centering
\includegraphics[width=0.65\linewidth]{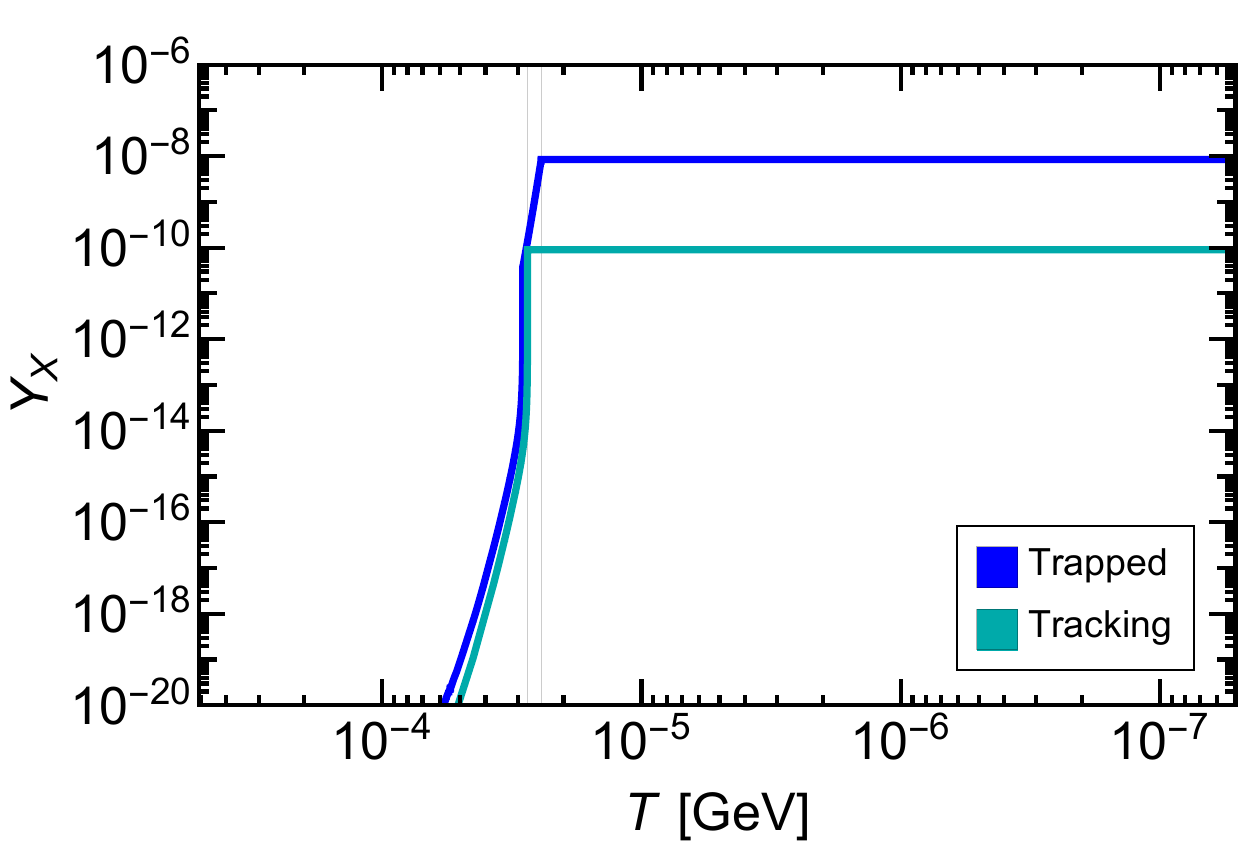}
\includegraphics[width=0.65\linewidth]{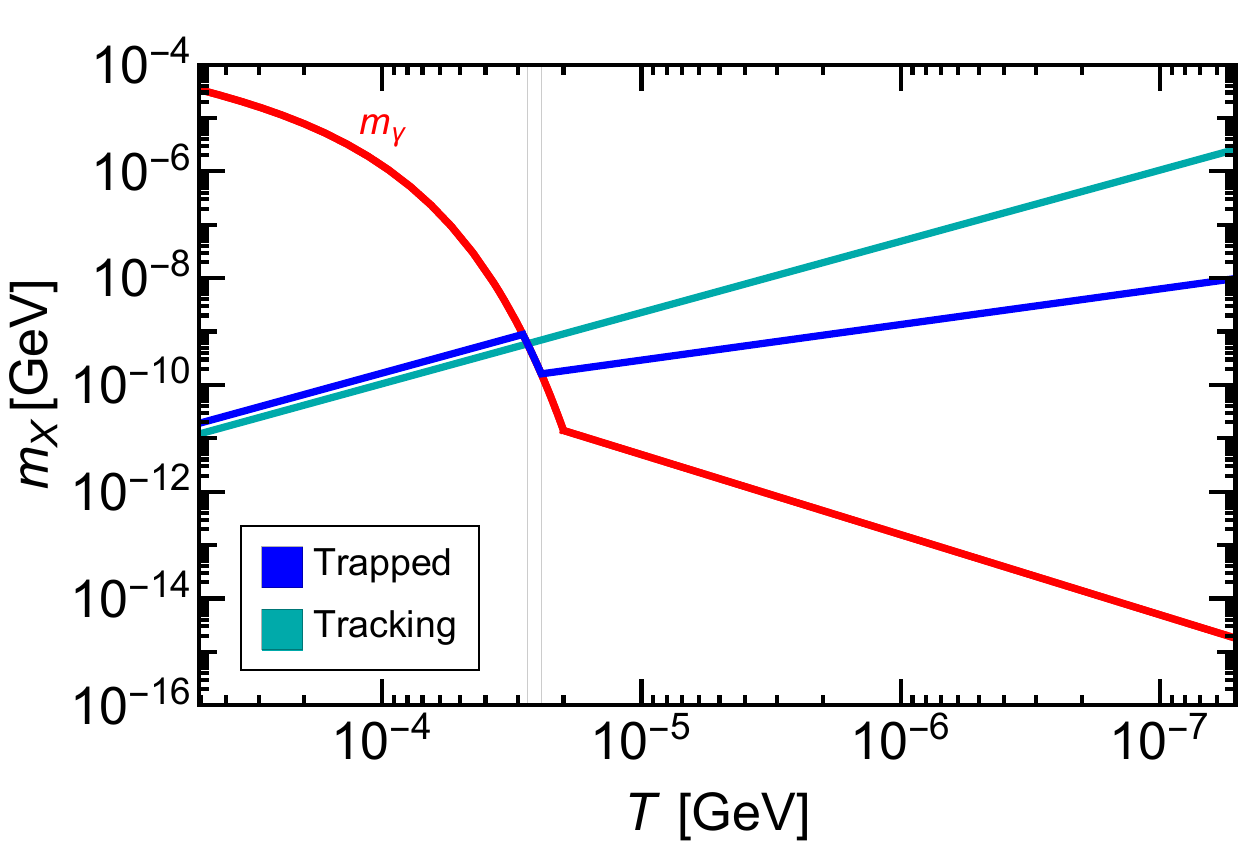}
\caption{Examples of the evolution of the dark gauge boson number density divided by entropy density, $Y_X \equiv n_X / s$, and the dark gauge boson mass $m_X$ around the time of resonant production.\cite{Kaneta:2023wdr} The blue curve corresponds to the case where the quintessence field is trapped in the potential minimum ($g_X = 5 \times 10^{-19}$), while the cyan curve represents the case where it is not trapped ($g_X = 3 \times 10^{-19}$). Both scenarios share a common kinetic mixing parameter, $\varepsilon = 10^{-11}$ and quintessence parameter $\alpha=1$. The thin vertical lines indicate when resonant production ends. In the figure below, the red curve denotes the effective photon mass $m_\gamma$. In the trapped case, the production is significantly enhanced compared to the tracking case, as the evolution of $m_X$ closely follows that of $m_\gamma$, effectively extending the resonance period.}
\label{fig:nkin}
\end{figure}

\subsection{Kinetic Mixing and Dark Photon Production}
\label{subsec:kinetic_mixing}

We introduce a kinetic mixing term between the dark gauge boson \(X\) (the ``dark photon'') and the SM photon \(\gamma\):\cite{Holdom:1985ag}
\begin{equation}
    \mathcal{L} \;\supset\; \frac{\varepsilon}{2}\,F_{\mu\nu}\,X^{\mu\nu},
\end{equation}
where \(F_{\mu\nu}\) and \(X_{\mu\nu}\) are the field strength tensors of the SM photon and dark photon, respectively, and \(\varepsilon\) is the (dimensionless) mixing parameter. Such a mixing permits non-gravitational interactions of \(X\) with the visible sector.

In a thermal plasma, the effective kinetic mixing \(\overline{\varepsilon}\) can differ from \(\varepsilon\). Following Ref.~\citenum{Redondo:2008aa}, the production rate in processes analogous to Compton scattering may be proportional to
\begin{equation}
    \bigl|\overline{\varepsilon}\bigr|^2 
    \;=\;
    \varepsilon^2\,\frac{m_X^4}{\bigl(m_X^2 - m_{\gamma}^2\bigr)^2 + \bigl(\omega D\bigr)^2},
\end{equation}
where \(m_\gamma\) is the effective photon plasma mass, \(\omega\) is the energy of the process, and \(D\) accounts for damping effects in the medium. This prescription allows dark photons to be produced in the early universe at rates depending on \(\varepsilon\), \(\omega\), \(m_X\), and the temperature-dependent plasma mass.

Since $|\bar{\epsilon}|^2$ resonantly increases when $m_X = m_\gamma$, the dark gauge boson is predominantly produced at this point. In models with a fixed mass, such as the conventional dark photon model, resonant production occurs only at a single moment in time. However, in the gauged quintessence model, the evolution of $m_X$ can track the evolution of $m_\gamma$ within certain regions of parameter space. As a result, the period of resonant production is extended over a finite time interval, allowing for the production of a larger number of dark gauge bosons. (See Fig. \ref{fig:nkin}.) 

\subsection{Mass-Varying Dark Photon Decays}
\label{subsec:dark_photon_decays}
In conventional dark photon models with a fixed mass \(m_X\), one typically encounters a single dominant decay channel. Below the \(e^+ e^-\) threshold (\(m_X < 2m_e\)), the decay proceeds via three photons, while above \(2m_e\) it proceeds primarily to \(e^+ e^-\). In gauged quintessence, however, \(m_X\) \emph{increases} over cosmic time, causing the dominant decay channel to \emph{change} at the epoch when \(m_X\) surpasses \(2m_e\). Consequently, dark photons may produce a characteristic time-dependent spectrum of photons and/or positron-electron pairs.

\begin{figure}[tb]
\centering
\includegraphics[width=0.8\linewidth]{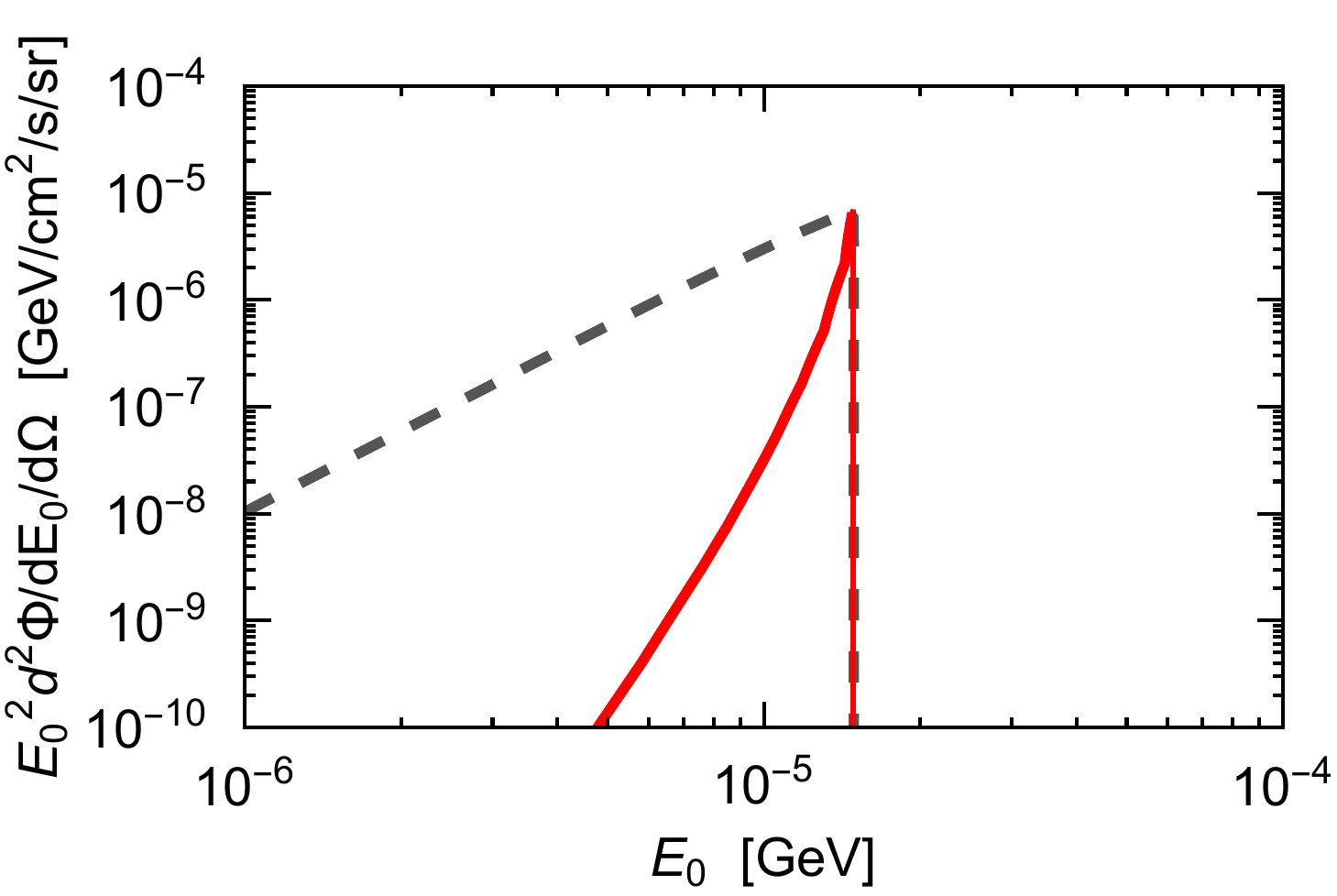}
\caption{Illustration of the diffuse $X$-ray/gamma-ray spectrum from the gauged quintessence (red) and conventional decaying dark matter model (dashed gray), highlighting the interplay between cosmic redshift and an increasing \(m_X(t)\).\cite{Kaneta:2023wdr}}
\label{fig:DiffSpec}
\end{figure}

\paragraph{Decay Channels.}
For \(m_X < 2m_e\), the decay rate into three photons (\(X \to 3\gamma\)) \cite{Pospelov:2008jk,McDermott:2017qcg} can be parametrized by
\begin{equation}
    \Gamma_{3\gamma} 
    = 
F(m_X)\frac{17\alpha_{\mathrm{em}}^4\overline{\varepsilon}^2}{11664000\pi^3}
    \frac{m_X^9}{m_e^8},
    \label{eq:3photondecayrate}
\end{equation}
where \(\alpha_{\mathrm{em}}\) is the electromagnetic fine-structure constant, \(m_e\) is the electron mass, and \(F(m_X)\) is an enhancement factor\cite{McDermott:2017qcg} that encodes the corrections at the mass threshold. Once \(m_X\) exceeds \(2m_e\), the dominant decay channel becomes \(X \to e^+ e^-\), with rate
\begin{equation}
    \Gamma_{e^+ e^-} 
    = \frac{\alpha_{\mathrm{em}}\overline{\varepsilon}^2\,m_X}{3}
    \sqrt{1 - \left(\frac{2m_e}{m_X}\right)^2}
    \left(1 + \frac{2\,m_e^2}{m_X^2}\right).
\end{equation}
Significantly, this implies that dark photons may yield a multi-stage decay history: they produce soft photons (when \(m_X\) is sub-\(2m_e\)) in earlier epochs and sharp positron–electron pairs once \(m_X>2m_e\). (See Fig.~\ref{fig:DiffSpec}).

\subsection{Observable Spectra and Timing}
\label{subsec:observable_signals}

Because \(m_X\) increases as \(\phi\) rolls to larger values and the decay rate scales with high powers of $m_X$, it increases rapidly as $m_X$ becomes larger. As a result, the high-energy region of the diffuse photon spectrum—originating from more recent decays—is enhanced relative to the low-energy region. This leads to a much steeper spectral slope compared to that predicted by conventional decaying dark matter models. Furthermore, once \(m_X\) surpasses \(1\,\mathrm{MeV}\), the decay \(X \to e^+ e^-\) can become almost immediate, producing non-relativistic positrons and electrons whose kinetic energy is correspondingly small.

\subsection{Constraints and Future Prospects}
\label{subsec:constraints_future}

\begin{figure}[tb]
\centering
\includegraphics[width=0.9\linewidth]{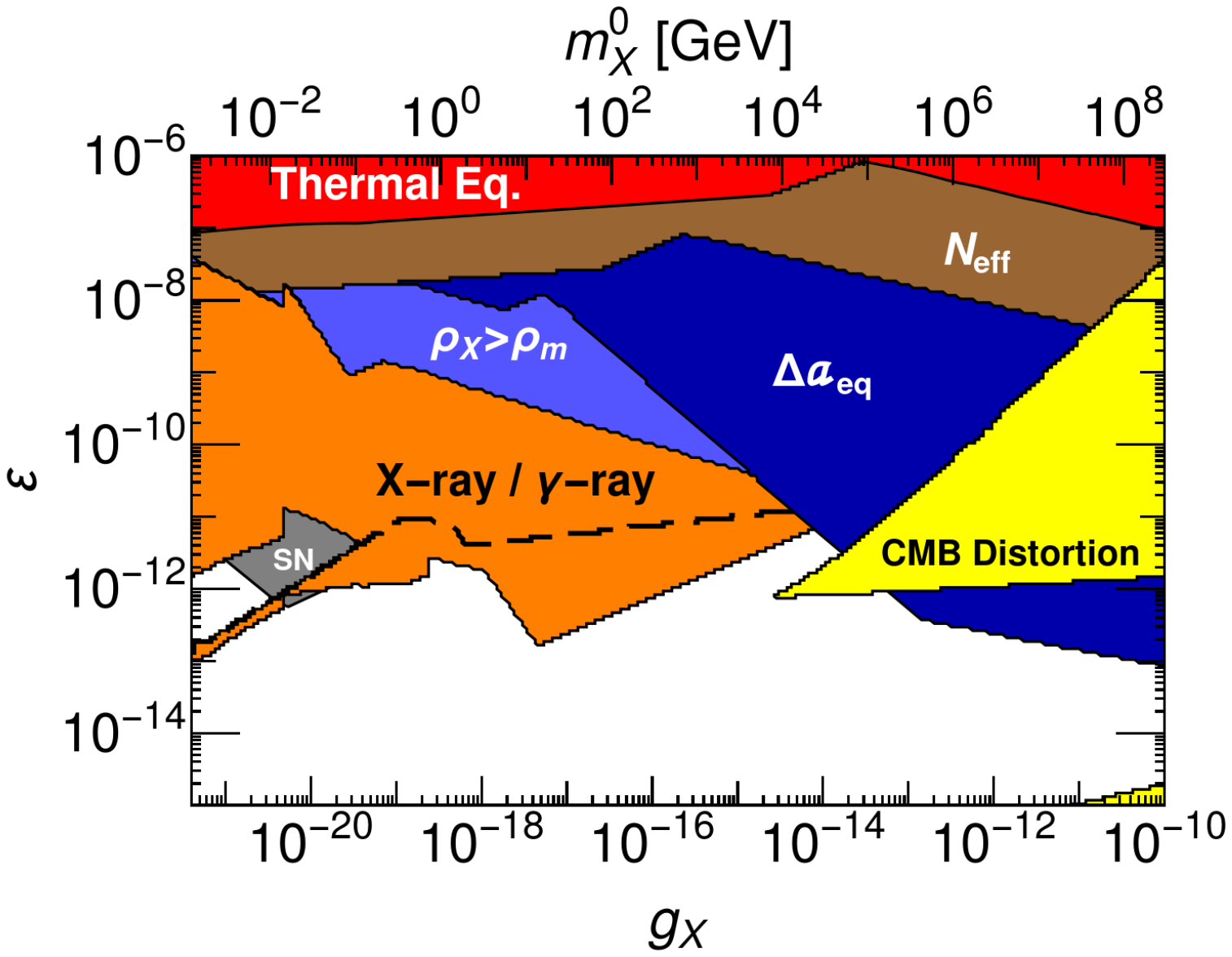}
\caption{A representative overview of constraints on the kinetic mixing parameter \(\varepsilon\) and the dark gauge coupling \(g_X\).\cite{Kaneta:2023wdr} The orange region is excluded by current X-ray/\(\gamma\)-ray diffuse background data. The blank region remains open to discovery, offering interesting prospects for future surveys.}
\label{fig:constraint2}
\end{figure}

Figure~\ref{fig:constraint2} summarizes current constraints on \((\varepsilon, g_X)\) from various astrophysical and cosmological observations, including \(\gamma\)-ray or \(X\)-ray surveys. In particular, the orange region denotes bounds derived from the diffuse photon background, which are sensitive to dark photon decays at various cosmic epochs. Future telescopes with improved sensitivity could potentially detect these novel signatures, thereby providing evidence of a time-varying dark photon mass.

In summary, gauged quintessence models allow the dark gauge boson \(X\) to carry a time-dependent mass. Once a kinetic mixing portal is introduced, these mass-varying dark photons can be produced or decayed at different epochs with shifting kinematics, thus imprinting unique signatures in various electromagnetic channels. Ongoing and next-generation observations of diffuse photon backgrounds, as well as high-precision \(X\)-ray/\(\gamma\)-ray instruments, can probe sizable portions of the \((\varepsilon, g_X)\) parameter space, opening a new window into the physics of dark energy beyond purely gravitational effects.

\section{Concluding Remarks}
\label{sec:conclusion}
In this review, we introduced the gauged extension of the well-known quintessence framework for dark energy. By coupling the quintessence scalar field \(\phi\) to a \(U(1)\) gauge boson \(X\), we demonstrated that the additional gauge potential \(V_\text{gauge}\) can significantly affect cosmological evolution and yield several novel phenomenological features. Among these are:

\begin{itemize}
    \item \textbf{Mass-Varying Vector Boson:} The rolling quintessence field endows the dark gauge boson with a time-dependent mass, potentially overcoming the usual suppression that hampers vector misalignment mechanisms.
    \item \textbf{Hubble Tension Alleviation:} Preliminary indications suggest that an interacting dark energy sector—especially one with an effective equation of state \(w_{\mathrm{eff}} < -1\) at late times—may help reconcile discrepancies between local and cosmological measurements of the Hubble constant. A quantitative study with precision data would be the next step in testing this possibility.
    \item \textbf{Gauge Principle in Dark Energy:} This work provides a proof of concept that the gauge principle, so instrumental in exploring the dark matter sector, can equally be applied to dark energy. Tools and techniques developed for analyzing dark matter symmetries are therefore directly relevant for investigations of gauged quintessence models.
\end{itemize}

Much like the ongoing, extensive efforts to uncover the nature of dark matter, the search for a governing symmetry in the dark energy sector remains open-ended. As this line of inquiry continues, more symmetries may come to light, and a fully unified model of the universe could ultimately integrate these new insights. We hope that gauged quintessence will serve as a useful paradigm for exploring physics beyond the minimal \(\Lambda\)CDM scenario, motivating future work in both theoretical and observational domains.

\vskip 0.2cm
\noindent \textbf{Acknowledgments.} 
The material in this review is based on Refs.~\cite{Kaneta:2022kjj,Kaneta:2023lki,Kaneta:2023wdr}.
This work was supported in part by the National Research Foundation of Korea (Grant No. RS-2024-00352537).

\bibliographystyle{ws-ijmpa}
\bibliography{sample}

\begin{thebibliography}{10}
\expandafter\ifx\csname urlstyle\endcsname\relax
  \providecommand{\doi}[1]{doi:\discretionary{}{}{}#1}\else
  \providecommand{\doi}{doi:\discretionary{}{}{}\begingroup \urlstyle{rm}\Url}\fi

\bibitem{Holdom:1985ag}
B.~Holdom, {\em Phys. Lett. B} {\bf 166}, 196  (1986), \doi{10.1016/0370-2693(86)91377-8}.

\bibitem{Peccei:1977hh}
R.~D. Peccei and H.~R. Quinn, {\em Phys. Rev. Lett.} {\bf 38}, 1440  (1977), \doi{10.1103/PhysRevLett.38.1440}.

\bibitem{Wilczek:1977pj}
F.~Wilczek, {\em Phys. Rev. Lett.} {\bf 40}, 279  (1978), \doi{10.1103/PhysRevLett.40.279}.

\bibitem{Weinberg:1977ma}
S.~Weinberg, {\em Phys. Rev. Lett.} {\bf 40}, 223  (1978), \doi{10.1103/PhysRevLett.40.223}.

\bibitem{Kim:1979if}
J.~E. Kim, {\em Phys. Rev. Lett.} {\bf 43},   103  (1979), \doi{10.1103/PhysRevLett.43.103}.

\bibitem{Shifman:1979if}
M.~A. Shifman, A.~I. Vainshtein and V.~I. Zakharov, {\em Nucl. Phys. B} {\bf 166}, 493  (1980), \doi{10.1016/0550-3213(80)90209-6}.

\bibitem{Zhitnitsky:1980tq}
A.~R. Zhitnitsky, {\em Sov. J. Nucl. Phys.} {\bf 31},   260  (1980).

\bibitem{Dine:1981rt}
M.~Dine, W.~Fischler and M.~Srednicki, {\em Phys. Lett. B} {\bf 104}, 199  (1981), \doi{10.1016/0370-2693(81)90590-6}.

\bibitem{Wess:1974tw}
J.~Wess and B.~Zumino, {\em Nucl. Phys. B} {\bf 70}, 39  (1974), \doi{10.1016/0550-3213(74)90355-1}.

\bibitem{Vagnozzi:2019kvw}
S.~Vagnozzi, L.~Visinelli, O.~Mena and D.~F. Mota, {\em Mon. Not. Roy. Astron. Soc.} {\bf 493}, 1139  (2020), \href{http://arxiv.org/abs/1911.12374}{{\ttfamily arXiv:1911.12374 [gr-qc]}}, \doi{10.1093/mnras/staa311}.

\bibitem{Ferlito:2022mok}
F.~Ferlito, S.~Vagnozzi, D.~F. Mota and M.~Baldi, {\em Mon. Not. Roy. Astron. Soc.} {\bf 512}, 1885  (2022), \href{http://arxiv.org/abs/2201.04528}{{\ttfamily arXiv:2201.04528 [astro-ph.CO]}}, \doi{10.1093/mnras/stac649}.

\bibitem{Vagnozzi:2021quy}
S.~Vagnozzi, L.~Visinelli, P.~Brax, A.-C. Davis and J.~Sakstein, {\em Phys. Rev. D} {\bf 104},   063023  (2021), \href{http://arxiv.org/abs/2103.15834}{{\ttfamily arXiv:2103.15834 [hep-ph]}}, \doi{10.1103/PhysRevD.104.063023}.

\bibitem{Kaneta:2022kjj}
K.~Kaneta, H.-S. Lee, J.~Lee and J.~Yi, {\em JCAP} {\bf 02},   005  (2023), \href{http://arxiv.org/abs/2208.09229}{{\ttfamily arXiv:2208.09229 [astro-ph.CO]}}, \doi{10.1088/1475-7516/2023/02/005}.

\bibitem{Kaneta:2023lki}
K.~Kaneta, H.-S. Lee, J.~Lee and J.~Yi, {\em JCAP} {\bf 09},   017  (2023), \href{http://arxiv.org/abs/2306.01291}{{\ttfamily arXiv:2306.01291 [astro-ph.CO]}}, \doi{10.1088/1475-7516/2023/09/017}.

\bibitem{Kaneta:2023wdr}
K.~Kaneta, H.-S. Lee, J.~Lee and J.~Yi, {\em JCAP} {\bf 03},   048  (2024), \href{http://arxiv.org/abs/2312.09717}{{\ttfamily arXiv:2312.09717 [astro-ph.CO]}}, \doi{10.1088/1475-7516/2024/03/048}.

\bibitem{Ratra:1987rm}
B.~Ratra and P.~J.~E. Peebles, {\em Phys. Rev. D} {\bf 37},   3406  (1988), \doi{10.1103/PhysRevD.37.3406}.

\bibitem{Peebles:1987ek}
P.~J.~E. Peebles and B.~Ratra, {\em Astrophys. J. Lett.} {\bf 325},   L17  (1988), \doi{10.1086/185100}.

\bibitem{Caldwell:1997ii}
R.~R. Caldwell, R.~Dave and P.~J. Steinhardt, {\em Phys. Rev. Lett.} {\bf 80}, 1582  (1998), \href{http://arxiv.org/abs/astro-ph/9708069}{{\ttfamily arXiv:astro-ph/9708069}}, \doi{10.1103/PhysRevLett.80.1582}.

\bibitem{Frieman:1995pm}
J.~A. Frieman, C.~T. Hill, A.~Stebbins and I.~Waga, {\em Phys. Rev. Lett.} {\bf 75}, 2077  (1995), \href{http://arxiv.org/abs/astro-ph/9505060}{{\ttfamily arXiv:astro-ph/9505060}}, \doi{10.1103/PhysRevLett.75.2077}.

\bibitem{Carroll:1998zi}
S.~M. Carroll, {\em Phys. Rev. Lett.} {\bf 81}, 3067  (1998), \href{http://arxiv.org/abs/astro-ph/9806099}{{\ttfamily arXiv:astro-ph/9806099}}, \doi{10.1103/PhysRevLett.81.3067}.

\bibitem{Kim:1998kx}
J.~E. Kim, {\em JHEP} {\bf 05},   022  (1999), \href{http://arxiv.org/abs/hep-ph/9811509}{{\ttfamily arXiv:hep-ph/9811509}}, \doi{10.1088/1126-6708/1999/05/022}.

\bibitem{Choi:1999xn}
K.~Choi, {\em Phys. Rev. D} {\bf 62},   043509  (2000), \href{http://arxiv.org/abs/hep-ph/9902292}{{\ttfamily arXiv:hep-ph/9902292}}, \doi{10.1103/PhysRevD.62.043509}.

\bibitem{Gu:2001tr}
J.-A. Gu and W.-Y.~P. Hwang, {\em Phys. Lett. B} {\bf 517}, 1  (2001), \href{http://arxiv.org/abs/astro-ph/0105099}{{\ttfamily arXiv:astro-ph/0105099}}, \doi{10.1016/S0370-2693(01)00975-3}.

\bibitem{Brisudova:2001wb}
M.~M. Brisudova, R.~P. Woodard and W.~H. Kinney, {\em Class. Quant. Grav.} {\bf 18}, 3929  (2001), \href{http://arxiv.org/abs/gr-qc/0105072}{{\ttfamily arXiv:gr-qc/0105072}}, \doi{10.1088/0264-9381/18/18/311}.

\bibitem{Boyle:2001du}
L.~A. Boyle, R.~R. Caldwell and M.~Kamionkowski, {\em Phys. Lett. B} {\bf 545}, 17  (2002), \href{http://arxiv.org/abs/astro-ph/0105318}{{\ttfamily arXiv:astro-ph/0105318}}, \doi{10.1016/S0370-2693(02)02590-X}.

\bibitem{Li:2001xaa}
X.-z. Li, J.-g. Hao and D.-j. Liu, {\em Class. Quant. Grav.} {\bf 19}, 6049  (2002), \href{http://arxiv.org/abs/astro-ph/0107171}{{\ttfamily arXiv:astro-ph/0107171}}, \doi{10.1088/0264-9381/19/23/311}.

\bibitem{Brisudova:2001ur}
M.~M. Brisudova, W.~H. Kinney and R.~P. Woodard, {\em Phys. Rev. D} {\bf 65},   103513  (2002), \href{http://arxiv.org/abs/hep-ph/0110174}{{\ttfamily arXiv:hep-ph/0110174}}, \doi{10.1103/PhysRevD.65.103513}.

\bibitem{Hill:2002kq}
C.~T. Hill and A.~K. Leibovich, {\em Phys. Rev. D} {\bf 66},   075010  (2002), \href{http://arxiv.org/abs/hep-ph/0205237}{{\ttfamily arXiv:hep-ph/0205237}}, \doi{10.1103/PhysRevD.66.075010}.

\bibitem{Kim:2002tq}
J.~E. Kim and H.~P. Nilles, {\em Phys. Lett. B} {\bf 553}, 1  (2003), \href{http://arxiv.org/abs/hep-ph/0210402}{{\ttfamily arXiv:hep-ph/0210402}}, \doi{10.1016/S0370-2693(02)03148-9}.

\bibitem{Mainini:2004he}
R.~Mainini and S.~A. Bonometto, {\em Phys. Rev. Lett.} {\bf 93},   121301  (2004), \href{http://arxiv.org/abs/astro-ph/0406114}{{\ttfamily arXiv:astro-ph/0406114}}, \doi{10.1103/PhysRevLett.93.121301}.

\bibitem{Rinaldi:2014yta}
M.~Rinaldi, {\em Class. Quant. Grav.} {\bf 32},   045002  (2015), \href{http://arxiv.org/abs/1404.0532}{{\ttfamily arXiv:1404.0532 [astro-ph.CO]}}, \doi{10.1088/0264-9381/32/4/045002}.

\bibitem{Mehrabi:2015lfa}
A.~Mehrabi, A.~Maleknejad and V.~Kamali, {\em Astrophys. Space Sci.} {\bf 362},  ~53  (2017), \href{http://arxiv.org/abs/1510.00838}{{\ttfamily arXiv:1510.00838 [astro-ph.CO]}}, \doi{10.1007/s10509-017-3033-z}.

\bibitem{Alvarez:2019ues}
M.~\'Alvarez, J.~B. Orjuela-Quintana, Y.~Rodriguez and C.~A. Valenzuela-Toledo, {\em Class. Quant. Grav.} {\bf 36},   195004  (2019), \href{http://arxiv.org/abs/1901.04624}{{\ttfamily arXiv:1901.04624 [gr-qc]}}, \doi{10.1088/1361-6382/ab3775}.

\bibitem{Orjuela-Quintana:2020klr}
J.~B. Orjuela-Quintana, M.~Alvarez, C.~A. Valenzuela-Toledo and Y.~Rodriguez, {\em JCAP} {\bf 10},   019  (2020), \href{http://arxiv.org/abs/2006.14016}{{\ttfamily arXiv:2006.14016 [gr-qc]}}, \doi{10.1088/1475-7516/2020/10/019}.

\bibitem{Motoa-Manzano:2020mwe}
J.~Motoa-Manzano, J.~Bayron Orjuela-Quintana, T.~S. Pereira and C.~A. Valenzuela-Toledo, {\em Phys. Dark Univ.} {\bf 32},   100806  (2021), \href{http://arxiv.org/abs/2012.09946}{{\ttfamily arXiv:2012.09946 [gr-qc]}}, \doi{10.1016/j.dark.2021.100806}.

\bibitem{Steinhardt:1999nw}
P.~J. Steinhardt, L.-M. Wang and I.~Zlatev, {\em Phys. Rev. D} {\bf 59},   123504  (1999), \href{http://arxiv.org/abs/astro-ph/9812313}{{\ttfamily arXiv:astro-ph/9812313}}, \doi{10.1103/PhysRevD.59.123504}.

\bibitem{Planck:2018vyg}
 Planck Collaboration (N.~Aghanim {\em et~al.}), {\em Astron. Astrophys.} {\bf 641},  ~A6  (2020), \href{http://arxiv.org/abs/1807.06209}{{\ttfamily arXiv:1807.06209 [astro-ph.CO]}}, \doi{10.1051/0004-6361/201833910}, [Erratum: Astron.Astrophys. 652, C4 (2021)].

\bibitem{Kolda:1998wq}
C.~F. Kolda and D.~H. Lyth, {\em Phys. Lett. B} {\bf 458}, 197  (1999), \href{http://arxiv.org/abs/hep-ph/9811375}{{\ttfamily arXiv:hep-ph/9811375}}, \doi{10.1016/S0370-2693(99)00657-7}.

\bibitem{Binetruy:1998rz}
P.~Binetruy, {\em Phys. Rev. D} {\bf 60},   063502  (1999), \href{http://arxiv.org/abs/hep-ph/9810553}{{\ttfamily arXiv:hep-ph/9810553}}, \doi{10.1103/PhysRevD.60.063502}.

\bibitem{Brax:1999gp}
P.~Brax and J.~Martin, {\em Phys. Lett. B} {\bf 468}, 40  (1999), \href{http://arxiv.org/abs/astro-ph/9905040}{{\ttfamily arXiv:astro-ph/9905040}}, \doi{10.1016/S0370-2693(99)01209-5}.

\bibitem{Doran:2002bc}
M.~Doran and J.~Jaeckel, {\em Phys. Rev. D} {\bf 66},   043519  (2002), \href{http://arxiv.org/abs/astro-ph/0203018}{{\ttfamily arXiv:astro-ph/0203018}}, \doi{10.1103/PhysRevD.66.043519}.

\bibitem{Coleman:1973jx}
S.~R. Coleman and E.~J. Weinberg, {\em Phys. Rev. D} {\bf 7}, 1888  (1973), \doi{10.1103/PhysRevD.7.1888}.

\bibitem{Jackiw:1974cv}
R.~Jackiw, {\em Phys. Rev. D} {\bf 9},   1686  (1974), \doi{10.1103/PhysRevD.9.1686}.

\bibitem{Preskill:1982cy}
J.~Preskill, M.~B. Wise and F.~Wilczek, {\em Phys. Lett. B} {\bf 120}, 127  (1983), \doi{10.1016/0370-2693(83)90637-8}.

\bibitem{Abbott:1982af}
L.~F. Abbott and P.~Sikivie, {\em Phys. Lett. B} {\bf 120}, 133  (1983), \doi{10.1016/0370-2693(83)90638-X}.

\bibitem{Dine:1982ah}
M.~Dine and W.~Fischler, {\em Phys. Lett. B} {\bf 120}, 137  (1983), \doi{10.1016/0370-2693(83)90639-1}.

\bibitem{Arias:2012az}
P.~Arias, D.~Cadamuro, M.~Goodsell, J.~Jaeckel, J.~Redondo and A.~Ringwald, {\em JCAP} {\bf 06},   013  (2012), \href{http://arxiv.org/abs/1201.5902}{{\ttfamily arXiv:1201.5902 [hep-ph]}}, \doi{10.1088/1475-7516/2012/06/013}.

\bibitem{Nelson:2011sf}
A.~E. Nelson and J.~Scholtz, {\em Phys. Rev. D} {\bf 84},   103501  (2011), \href{http://arxiv.org/abs/1105.2812}{{\ttfamily arXiv:1105.2812 [hep-ph]}}, \doi{10.1103/PhysRevD.84.103501}.

\bibitem{Riess:2021jrx}
A.~G. Riess {\em et~al.}, {\em Astrophys. J. Lett.} {\bf 934},  ~L7  (2022), \href{http://arxiv.org/abs/2112.04510}{{\ttfamily arXiv:2112.04510 [astro-ph.CO]}}, \doi{10.3847/2041-8213/ac5c5b}.

\bibitem{DiValentino:2016hlg}
E.~Di~Valentino, A.~Melchiorri and J.~Silk, {\em Phys. Lett. B} {\bf 761}, 242  (2016), \href{http://arxiv.org/abs/1606.00634}{{\ttfamily arXiv:1606.00634 [astro-ph.CO]}}, \doi{10.1016/j.physletb.2016.08.043}.

\bibitem{DiValentino:2017iww}
E.~Di~Valentino, A.~Melchiorri and O.~Mena, {\em Phys. Rev. D} {\bf 96},   043503  (2017), \href{http://arxiv.org/abs/1704.08342}{{\ttfamily arXiv:1704.08342 [astro-ph.CO]}}, \doi{10.1103/PhysRevD.96.043503}.

\bibitem{Joudaki:2017zhq}
S.~Joudaki, M.~Kaplinghat, R.~Keeley and D.~Kirkby, {\em Phys. Rev. D} {\bf 97},   123501  (2018), \href{http://arxiv.org/abs/1710.04236}{{\ttfamily arXiv:1710.04236 [astro-ph.CO]}}, \doi{10.1103/PhysRevD.97.123501}.

\bibitem{Heisenberg:2022lob}
L.~Heisenberg, H.~Villarrubia-Rojo and J.~Zosso, {\em Phys. Dark Univ.} {\bf 39},   101163  (2023), \href{http://arxiv.org/abs/2201.11623}{{\ttfamily arXiv:2201.11623 [astro-ph.CO]}}, \doi{10.1016/j.dark.2022.101163}.

\bibitem{Heisenberg:2022gqk}
L.~Heisenberg, H.~Villarrubia-Rojo and J.~Zosso, {\em Phys. Rev. D} {\bf 106},   043503  (2022), \href{http://arxiv.org/abs/2202.01202}{{\ttfamily arXiv:2202.01202 [astro-ph.CO]}}, \doi{10.1103/PhysRevD.106.043503}.

\bibitem{Lee:2022cyh}
B.-H. Lee, W.~Lee, E.~O. Colg\'ain, M.~M. Sheikh-Jabbari and S.~Thakur, {\em JCAP} {\bf 04},   004  (2022), \href{http://arxiv.org/abs/2202.03906}{{\ttfamily arXiv:2202.03906 [astro-ph.CO]}}, \doi{10.1088/1475-7516/2022/04/004}.

\bibitem{Banerjee:2020xcn}
A.~Banerjee, H.~Cai, L.~Heisenberg, E.~O. Colg\'ain, M.~M. Sheikh-Jabbari and T.~Yang, {\em Phys. Rev. D} {\bf 103},   L081305  (2021), \href{http://arxiv.org/abs/2006.00244}{{\ttfamily arXiv:2006.00244 [astro-ph.CO]}}, \doi{10.1103/PhysRevD.103.L081305}.

\bibitem{baoimg}
Z.~Rostomian. \url{https://newscenter.lbl.gov/wp-content/uploads/BOSS-BAO.jpg},  (2014).

\bibitem{Essig:2013lka}
R.~Essig {\em et~al.}, { {Working Group Report: New Light Weakly Coupled Particles}}, in {\em {Snowmass 2013}: {Snowmass on the Mississippi}\/},  (10 2013), \href{http://arxiv.org/abs/1311.0029}{{\ttfamily arXiv:1311.0029 [hep-ph]}}.

\bibitem{Redondo:2008aa}
J.~Redondo, {\em JCAP} {\bf 07},   008  (2008), \href{http://arxiv.org/abs/0801.1527}{{\ttfamily arXiv:0801.1527 [hep-ph]}}, \doi{10.1088/1475-7516/2008/07/008}.

\bibitem{Pospelov:2008jk}
M.~Pospelov, A.~Ritz and M.~B. Voloshin, {\em Phys. Rev. D} {\bf 78},   115012  (2008), \href{http://arxiv.org/abs/0807.3279}{{\ttfamily arXiv:0807.3279 [hep-ph]}}, \doi{10.1103/PhysRevD.78.115012}.

\bibitem{McDermott:2017qcg}
S.~D. McDermott, H.~H. Patel and H.~Ramani, {\em Phys. Rev. D} {\bf 97},   073005  (2018), \href{http://arxiv.org/abs/1705.00619}{{\ttfamily arXiv:1705.00619 [hep-ph]}}, \doi{10.1103/PhysRevD.97.073005}.

\end{thebibliography}

\end{document}